\documentclass[a4paper0,12pt]{article}
\usepackage[centertags]{amsmath}
\usepackage{amsfonts}
\usepackage{graphicx}
\usepackage{amsbsy}
\usepackage{cite}

\begin{document}

\setlength{\unitlength}{1cm}
\newcommand{\be}{\begin{eqnarray}}
\newcommand{\ee}{\end{eqnarray}}
\newcommand{\bee}{\begin{eqnarray*}}
\newcommand{\eee}{\end{eqnarray*}}
\newcommand{\pmat} {\begin{array}}
\newcommand{\ep} {\end{array}}
\newcommand{\ra}{\rightarrow}
\newcommand{\sse}{\subsection}
\newcommand{\+} {\`}
\newcommand{\es} {\epsilon}
\newcommand{\pa} {\partial}
\newcommand{\ddh} {\dd h}
\newcommand{\al} {\alpha}
\newcommand{\dd} {\dot}
\newcommand{\hb} {\hbar}
\newcommand{\R}{\mbox {\sc R}}
\newcommand{\N}{\mbox {\sc N}}
\newcommand{\Z}{\mbox {\sc Z}}
\newcommand{\C}{\mbox {\sc C}}
\newcommand{\D}{\mbox {\sc D}}
\newcommand{\I}{\mbox {\sc 1}}
\newcommand{\T}{\mbox {\sc T}}
\newcommand{\s}{\mbox {\sc P}}
\newcommand{\p}{\mbox {\sc T}}
\newcommand{\0}{\mbox {\sc 0}}
\newcommand{\Rp}{\mbox {\sc r}}
\newcommand{\Np}{\mbox {\sc n}}
\newcommand{\Zp}{\mbox {\sc z}}
\newcommand{\Cp}{\mbox {\sc c}}
\newcommand{\asy}{{\cal O}

\newcommand{\ind}{\hskip 0.5cm}}

\def\spazio#1{\vrule height#1em width0em depth#1em}

\title{ Level Crossings\\ in a \\\textit{PT}-symmetric Double Well}
\author{\textbf{Riccardo Giachetti}\\Dipartimento di Fisica e Astronomia, Universit\+a di 
	Firenze,\\ 50019 Sesto Fiorentino.  I.N.F.N. Sezione di Firenze. \\\\ \textbf{Vincenzo 
		Grecchi}\\Dipartimento di Matematica, Universit\+a di Bologna, 40126 Bologna.\\ I.N.F.N. 
	Sezione di Bologna.}
\maketitle
\textbf{Abstract}\\
\small{We consider a  \textit{PT}-symmetric cubic oscillator with an imaginary 
double well.   	We prove the existence of an infinite number of level crossings 
with a definite selection rule. 
Decreasing the
positive parameter $\hb$ from large values, at a  parameter $\hb_n$ we find the crossing of the 
pair of levels $(E_{2n+1}(\hb),E_{2n}(\hb))$ becoming the pair of levels    
$(E_n^+(\hb),E_n^-(\hb))$. For  large parameters,  a level is a  holomorphic function  
$E_m(\hb)$  with   different semiclassical  behaviors, $E_j^\pm(\hb),$  along different paths.  
The corresponding $m$-nodes delocalized  state $\psi_m(\hb)$  behaves along the same paths as 
the semiclassical $j$-nodes states $\psi_j^\pm(\hb),$  localized at one of the wells  $x_\pm$ 
respectively.     In particular, if    
the crossing parameter  $\hb_n$ is by-passed from above, the  levels $E_{2n+(1/2)\pm(1/2)}(\hb)$  have respectively  
the semiclassical behaviors of the levels $E_n^\mp(\hb)$ along the real axis.   These results are obtained by the 
control of the  nodes. There is evidence that the parameters $\hb_n$  accumulate at zero and the 
accumulation 
point of the corresponding energies is 	an	instability point of a subset of the Stokes complex 
called the monochord, consisting of the vibrating string and the sound board.}
\section{Introduction}
\bigskip 

The anharmonic oscillators are among the simplest non solvable models in quantum mechanics. In 
addition to presenting some connections with quantum field theory, their main interest mainly lies 
in the presence of diverging perturbation series and in the problem of their summability 
\cite{LM}.  
The latter is related to the existence of singularities of the levels as functions of the 
perturbation parameter. Since the family of Hamiltonians is analytic \cite{K}, 
such singularities are due to level crossings.  The semiclassical theory provides good qualitative 
and quantitative results for lower parameter up to the crossing value 
\cite{BW,BG, BOG,A}.
The  exact semiclassical method \cite{DDP} has extended the results to 
larger, but not very large, values of the parameter \cite{DP,DT}. Such results are useful 
and complementary to the rigorous results we are showing here. Indeed we believe
that only the nodal 
analysis, begun in the papers \cite{SHA,E,EGS,GM}, can give  a clear and 
exhaustive analysis of the level crossings, for which  a generalization of the method of 
control of the zeros by Loeffel-Martin \cite{LM} 
as well as a generalized semiclassical theory are useful tools.
Unfortunately, it is not  easy to prove the existence  of these crossings, and it is even  
harder to give the selection rule on the two pair of levels (at least) involved in a crossing.
The first problem is the unique labeling of the levels.
Rigorous results were recently obtained  in \cite{EG} by different techniques.
The present paper was announced in \cite{GG} and its purpose, as we said, is to produce  
rigorous results by a clear method based on nodal analysis and and making recourse to some physical
 notions. We will  also make some hypotheses in order to extend 
the treatment and exhibit a complete understanding of the full phenomenon.

Level crossings are forbidden in case of analytic families of self-adjoint
Hamiltonians \cite{K}; also in the case of families of single well Hamiltonians with 
\textit{PT}-symmetry
\cite{BG,BOG,BB}  the absence of crossings was proved in  \cite{SH,C}.  Andr\'e Martinez and one of 
us
(V. G.) in the paper  \cite{GM} have extended the proof of absence of crossings  of the 
perturbative levels $\tilde{E}_n(\beta)$, $n\in\mathbb{N},$ of  the analytic family of single well 
cubic 
oscillators (\ref{HB2}),
$$ H(\beta)=p^2+x^2+i\sqrt{\beta}x^3,\,\quad p^2=-\frac{d^2}{dx^2},\,\quad\beta\neq 0, 
\,\quad |\arg(\beta)|<\pi\,$$   
with fixed domain $D(H(\beta))=D(p^2)\bigcap D(|x|^3)$. The  labeling of the states 
$\tilde{\psi}_n(\beta)$ and the  corresponding levels $\tilde{E}_n(\beta)$ is based on  the  $n$ 
nodes as the  stable zeros at $\beta=0$.
In \cite{GM}  the semiclassical method was also used, but the exact results of 
analyticity were mostly given by the control of the nodes of the states.   
Our program is to extend the 
analysis of the perturbative levels to the other regions of $\beta$ where  the complex potential 
presents a double well structure and where the existence of crossings is expected. 
In the case of \textit{PT}-symmetric double wells we expect to have level crossings for real 
$\hb$.
We then continue the Hamiltonian $H(\beta)$  to the two sectors 
$\pi<|\arg(\beta)|<3\pi/2,$ by using the complex dilations. 
By two possible  changes of 
representation in the extended sense, with the parameter  
transformations 
\be \beta^\pm(\hb)=\exp(\mp i5\pi/4) 3^{-5/4}\hb,\label{TBH}\ee 
for $\pi<\mp \arg(\beta)<3\pi/2$ respectively (\ref{dila},\ref{PSR1}),
we get 
the semiclassical family of Hamiltonians
\be\,\,\,H_\hb=\hb^2p^2+V(x),\,\,V(x)=i(x^3- x),\,\,\, \hb>0, \,\,\,\,\,\,\,\label{RKHKC1}\ee 
on the same domain $D(p^2)\bigcap D(|x|^3)$ for
\be\hb\in\mathbb{C}^0=\{\hb\in \mathbb{C}; \,\hb\neq 0, \,|\arg(\hb)|<\pi/4\}.\label{SIH}\ee
The Hamiltonians $H_\hb$ for $\hb>0$ are closed and 
\textit{PT}-symmetric operators.  Since the derivative of the potential $V'(x)$ has  two  real 
zeros at $x_\pm=\pm 1/\sqrt{3}$, $H_\hb$ can be regarded as a double well Hamiltonian: it is indeed 
a peculiar  double well without an internal barrier and we will see that for 
complex energy it is, actually, an 
effective single well Hamiltonian.

Let us recall something about the real double wells.
As a simple example, we consider a self-adjoint Hamiltonian with a double well potential as 
$V(x)=(x^2-1)^2$.  For $E>V(0)=1$ we have a semiclassical regime of delocalized states, and for 
$E<1$ we have a semiclassical regime of bilocalized states. Localized states in a single well can 
exist for complex $\hb$.  We expect the existence of level crossings for almost real parameters 
$\hb$ and energies near the critical energy given by the internal top of the potential, $V(0)=1.$
One indication of this fact comes from the presence of a logarithmic 
term in the separation distance of the levels \cite{HS}. 

 Coming back to our case, the critical energy can be defined by studying the Stokes complex. 
 Since we know the absence  of singularities of the level $E_n(\hb)$ for 
 small  
 $|\hb|$  in certain sectors \cite{C}, we define two other types of levels  for 
 small $\hb>0,$ by the analytic continuations of $\hat{E}_n(\al)$ on the complex plane along  arcs 
 of circle of radius  $|\al|$, starting from $\al=\hb^{-4/5}>0$  and arriving to 
 $\al^\pm:=\exp(\pm i\pi)\al$, respectively.
 We thus define the levels
 \be E_n^\pm(\hb):=\hb^{6/5}\hat{E}_n(\al^\mp), \,\,n\in\mathbb{N},\,\,\,\al^\pm=\exp(\pm 
 i\pi)\hb^{-4/5},\,\,\,\hb>0.\,\,\label{RHT1}\ee
 All such levels $E_n^\pm(\hb)$ are  analytic continuations of the perturbative levels $\tilde 
 E_n(\beta)$ as $\tilde E_n(\beta^\pm(\hb))$ by the relations  (\ref{TBH}) and are extensible as 
 many-valued functions to the sector $\mathbb{C}^0$ of the $\hb$ complex plane.
The large $\hb$ behavior of the level
${E}_m(\hb)$  is studied using a different scaling that gives a new representation with 
the  Hamiltonians
\be K(\al)=p^2+W(\al, x),\,\,\,W(\al,x)=i(x^3+\al x),\,\,\,\,\,\al\in\mathbb{C}.\label{KH}\ee
    The level $\hat{E}_m(\al)$,  of $K(\al)$ is holomorphic on the sector,
\be\mathbb{C}_\al=\{\al\in\mathbb{C},\, \al\neq 0, |\arg(\al)|< 4\pi/5\},\label{ST}\ee
but before 
the first crossing it can be analytically continued \cite{GM} as a positive function up to negative 
values  by the relation,
\be E_m(\hb)=\hb^{6/5}\hat{E}_m(\al),\qquad \al=-\hb^{-4/5}\,\,.\label{RHT}\ee
The levels $E_m(\hb)$ for large $\hb$ are related to the perturbative levels by  
(\ref{RHT}) and the behavior (\ref{AIB}),$$\tilde E_m(\beta)\sim \beta^{1/5}\hat E_m(0)\,\,\, 
\textrm{as} \,\,\,\beta\ra+\infty.$$
All such levels $E_m(\hb)$ are  analytic continuations of the perturbative levels  $\tilde 
E_m(\beta)$ and
are extensible as many-valued functions, to the sector $\mathbb{C}^0$ of the $\hb$ complex plane. 

We give all the rules of the crossings in a minimality hypothesis which allows   to simplify the 
notations.

\bigskip 
{\textsc{Hypothesis H1}} \textit{The number of crossings involving two given pairs of levels, 
respectively before and after the crossing, is minimal.} 

\bigskip
As a result, the crossing parameter $\hbar_n$ is unique. 
Both the levels $E_n^\pm(\hb),\,$ such that
\be E_n^+(\hb)=\bar{E}_n^-(\hb), \label{CCPM}\ee
are non-real analytic for small $\hb>0$ 
and have the  semiclassical behaviors 
(\ref{0B}),
\be E_n^\pm(\hb)=\mp i\frac{2}{3\sqrt{3}}+ \sqrt{\pm 
	i}\sqrt[4]{3}(2n+1)\,\hb+O(\hb^{2})\,\in\,\mathbb{C}_\mp =\{z\in\mathbb{C},\, \mp \Im 
	z>0\}.\label{CPM1}\ee
Since all the levels are real for large $\hb>0$, there exist $\hb_n>0$ such that the levels 
$E^\pm(\hb_n)$ 
are real and equal because of (\ref{CCPM}). Thus, the first part of the crossing rule (Theorem 1) 
is proved.
For the second part,
at a fixed parameter  $\hb>0$, we extend  the states $\psi_n^\pm (x) $ and the state  
$\psi_m(x)$, 
 $x\in\mathbb{R}$,  as entire functions on the complex $z=x+iy$ plane. 
In particular,  the state $\psi_m(z)$ corresponding to a positive level $E_m$  is taken to be  
$P_xT$-symmetric, where $$P_x\psi_m(x+iy)=\psi_m(-x+iy).$$ 
We now prove  that  for $\hb<\hb_n$ the $n$ nodes of  $\psi_n^\pm(\hb)$ are their only 
zeros in 
\be\mathbb{C}^\pm:=\{z\in\mathbb{C},\, \pm \Re z>0\},\label{CPM}\ee 
respectively.   
At the left limit, $\hb\ra \hb_n^-,$  the union of the sets of $n$ nodes of the two states 
$\psi_n^\pm(\hb)$ becomes the $P_x$-symmetric set of $2n$ non imaginary zeros of the critical state 
$\psi_{n,n}^c$  (Lemma 7). The state $\psi=\psi_{n,n}^c$ is completely $P$-asymmetric in the sense 
that the mean parity vanishes, $\langle\psi,P \psi\rangle=0$ (Lemma 9). 
At the right limit  $\hb\ra \hb_n^+,$
the sets of $2n$ non imaginary zeros of both  the new states, generically called $\psi_m(\hb)$,   
are stable (Lemma 8).
In Theorem 1  we show that for $\hb>\hbar_n$ all the non imaginary  zeros of the states 
$\psi_m(\hb)$  are locally stable. The label $m$ is the number of zeros (nodes) of the state 
$\psi_m$ in $\mathbb{C}_-$ for large $\hb>0$  (Lemma 1).  The number of non imaginary nodes can be 
$2j$, 
$0\leq j\leq n.$
It is possible that no imaginary node or only one imaginary node does exist (Lemma 4). 
Thus $2n+1$ is the maximum value $m.$ Since the two values of $m$ must be different for the 
independence of the states, the maximum values of the pair of integer $m$ is $(2n,2n+1)$. If we 
consider the sequence of levels obtained by the crossings, the sequence of the maximum values is 
the only one compatible with the 
 uniqueness of the   state $\psi_m$ for a given $m$.
Therefore, for $\hb>\hb_n$, the pair of independent states $\psi_m(\hb)$,  continuation of the 
pair of states $\psi_n^\pm(\hb)$, is   
$(\psi_{2n}(\hb),\,\psi_{2n+1}(\hb))$ corresponding to the pair of levels $(E_{2n}(\hb),\, 
E_{2n+1}(\hb))$. Only the state $\psi_{2n+1}(\hb)$ has an imaginary node. 
The levels are locally bounded as proved in Lemma 10.   

The crossing selection rule can also be given in simple terms. The two  levels  
$E_{2n+(1/2)\pm (1/2)}(\hb)$,  separated for $\hb>\hbar_n,$ 
cross at $\hbar_n>0$, becoming the two separated levels $E_n^\pm(\hb)$ for $\hb<\hbar_n.$
We call $E^c_{n,n}$  the limit level at $\hb=\hb_n$ and $\psi^c_{n,n}$ the corresponding state.  
More explicitly, 
the crossing rule is given in terms of the analytic continuations  (Theorem 2). 
The two functions $\,E_{2n+(1/2)\pm (1/2)}(\hb)\,$, holomorphic for large $\,|\hb|\,$, 
are analytically continued along the positive semi-axis for decreasing $\hb>0$  by passing above 
the singularity at $\hb=\hbar_n$ as,
for instance, along a semi-circle of radius $\es>0$ and parameter $\theta$
$$\hb(\theta)-\hbar_n=\es\exp( i\theta),\,\,\,\,\theta\in[0,\pi],$$ 
They have respectively the two semiclassical  behaviors $E_n^\mp(\hb)$ for small  
positive $\hb$.  

All the results presented so far have been rigorously proved. We now continue our 
investigation introducing some definitions and making some conjectures,
arising on the basis of numerical results, that we believe 
useful for a full understanding of this specific problem.
\bigskip

\textbf{Definitions}.
Let $\hb=0$.  We call (\textit{vibrating})  \textit{string} the short Stokes line  \cite{EGS}, (\textit{sound}) 
\textit{board} 
the exceptional Stokes line  \cite{EGS}. Their union is a subset of the Stokes complex called the 
\textit{monochord}.
\bigskip

{\textsc{Conjecture C1}} \textit{Let us fix $\hb\in\mathbb{C}^0$, and consider the state  $\psi_E$ 
corresponding to the level $E.$ There exists  the string, an arc of line where the nodes lie, and 
the board, a half-line where the other zeros lie. 
The string is the exact short Stokes line and the board is the exact exceptional Stokes line in 
the sense of the exact semiclassical theory $(\ref{EXACTSTOKES})$, {\rm{\cite{DDP}}}. The 
approximate monochord 
is exact at $\hb=0,$ and the approximated board is exact in case of a positive level $E_m(\hb)$ 
at a positive parameter $\hb$.} 

\bigskip
These notions are relevant in order to  control the stability of the nodes for any $\hb$.
The numerical results, reported in Fig. 4-7, support the conjecture \textsc{C1}.
A node can disappear
by passing  from the string to the board. On the other side an antinode can double 
after a crossing with 
a stationary point at a turning point.  
These events are  possible when the string and the board come in contact.
\begin{figure}
	\centerline{\includegraphics[height=180pt]{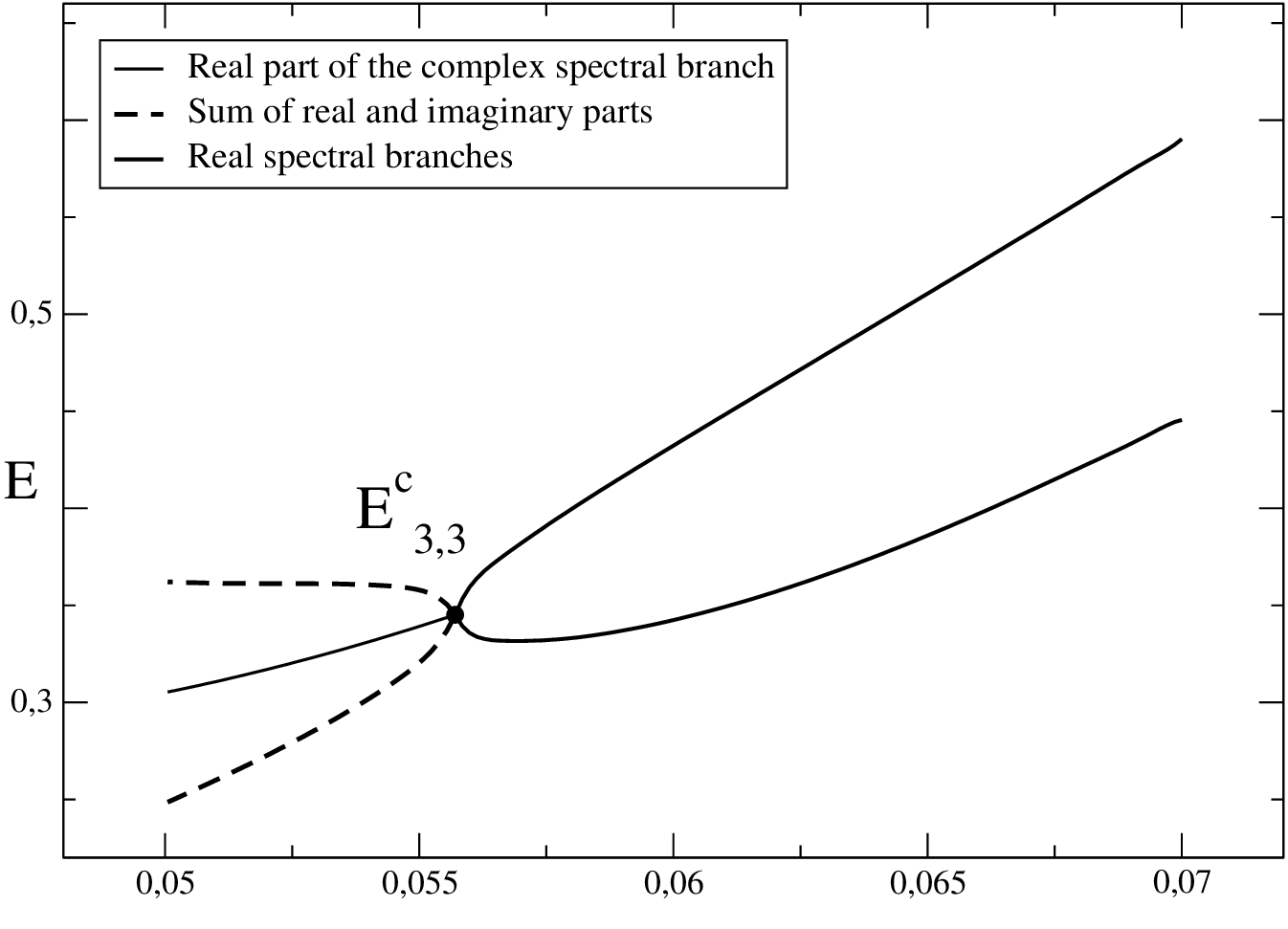}}
	\caption{The crossing process of the pair of levels $(E_6(\hb),E_7(\hb))$ for $\hb>h_3$, and 
		the pair of levels $E^\pm_3(\hb)$ for $\hb<h_3$. A complex level $E$ is represented by 
		$\Re E+\Im E$.}
	\label{Fagioli}	
\end{figure}

\bigskip
 
{\textsc{Conjecture C2}} \textit{The standard sequence of nodes and antinodes  of a state 
$\psi_m(\hb)$  with m nodes, $[m/2]=n$, for  $\hb\gg\hbar_n>0$ and suitable labeling,  is the 
following:
$$S_{2n}=(A_{-n-1}, ~N_{-n},\,\,...\,\,,A_{-2},~N_{-1},~A_0,~N_1,~A_2,\,\,....\,\,,N_n,~A_{n+1}),$$
$$S_{2n+1}=(A_{-n-1}, ~N_{-n},\,\,...\,\,,A_{-1},~N_0,~A_1,\,\,....\,\,,N_n,~A_{n+1}).$$
There exists a  parameter $\hbar_n^a>\hbar_n$ such that the antinode $A_0$ of the state 
$\psi_{2n}(h^a_n)$ coincides with the 
imaginary turning point  $I_0$ {\rm{(Remark 1)}}. 
There exists a  parameter $\hbar_n^p>\hbar_n$ such that the node 
$N_0$ of the state $\psi_{2n+1}(h^p_n)$ coincides with the imaginary turning point  $I_0$ 
{\rm{(Lemma 4)}}. This 
means that at the  parameter $\hbar_n^a$ and energy  $E_{2n}(\hb^a_n)$ the end point of the 
board, $I_0$, touches the string. The same happens at the parameter $\hbar_n^p$ and energy 
$E_{2n+1}(\hb^p_n)$.
Decreasing $\hb,$ just below $\hb^a_n$ the imaginary antinode $A_0$ of $S_{2n}$ doubles into the 
pair 
of non imaginary antinodes $(A_{-1},\,A_{1})$, and just below $\hb^p_n$ the imaginary node $N_0$ of 
$S_{2n+1}$ disappears.
Thus, the sequence of nodes of the state at the crossing, $\psi_{n,n}^c$, is, 
$$S^c_{n,n}=(A_{-n-1},~N_{-n},\,\,...\,\,,N_{-1},~A_{-1},~A_1,~N_1,....\,\,,N_n,~A_{n+1}),$$ 
and the sequences of the 
nodes of the states $\psi_n^\pm(\hb)$ for $\hb<\hbar_n$ are 
$$ S^-_n=(A_{-n-1},~N_{-n},\,\,...\,\,,N_{-1}, ~A_{-1}),\qquad 
S^+_n=(A_1,~N_1,\,\,...\,\,,N_{n},~A_{n+1}) 
$$ 
respectively.}
\bigskip\bigskip

\begin{figure}[h]
	\centerline{\includegraphics[height=200pt]{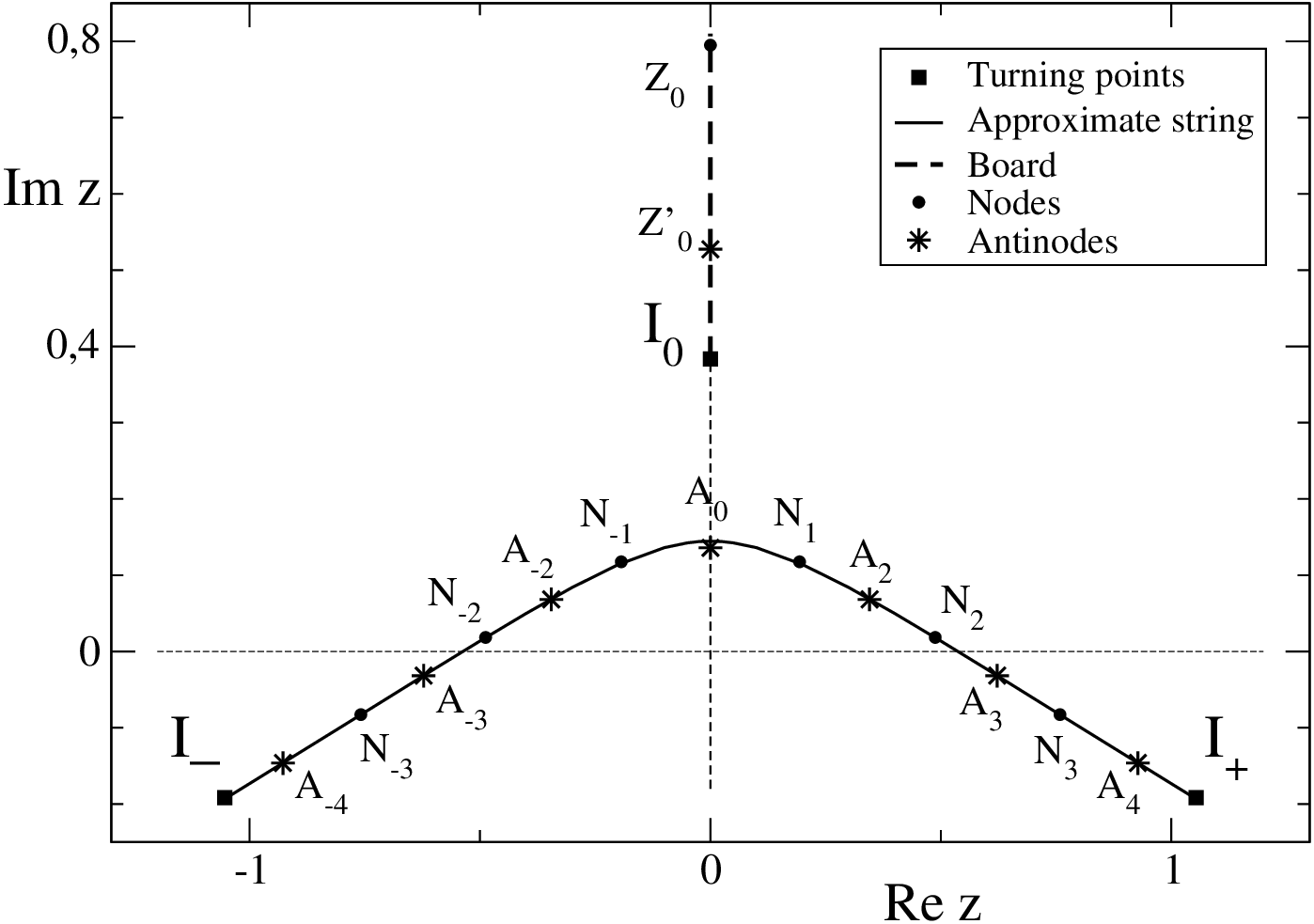}}
	\caption{ $\hb>h_3^a$. The approximate monochord of $E_{6}(\hb)$  with the nodes and 
		antinodes.}
	\label{Nodi_Antinodi_044}	
\end{figure}
\begin{figure}
	\centerline{\includegraphics[height=200pt]{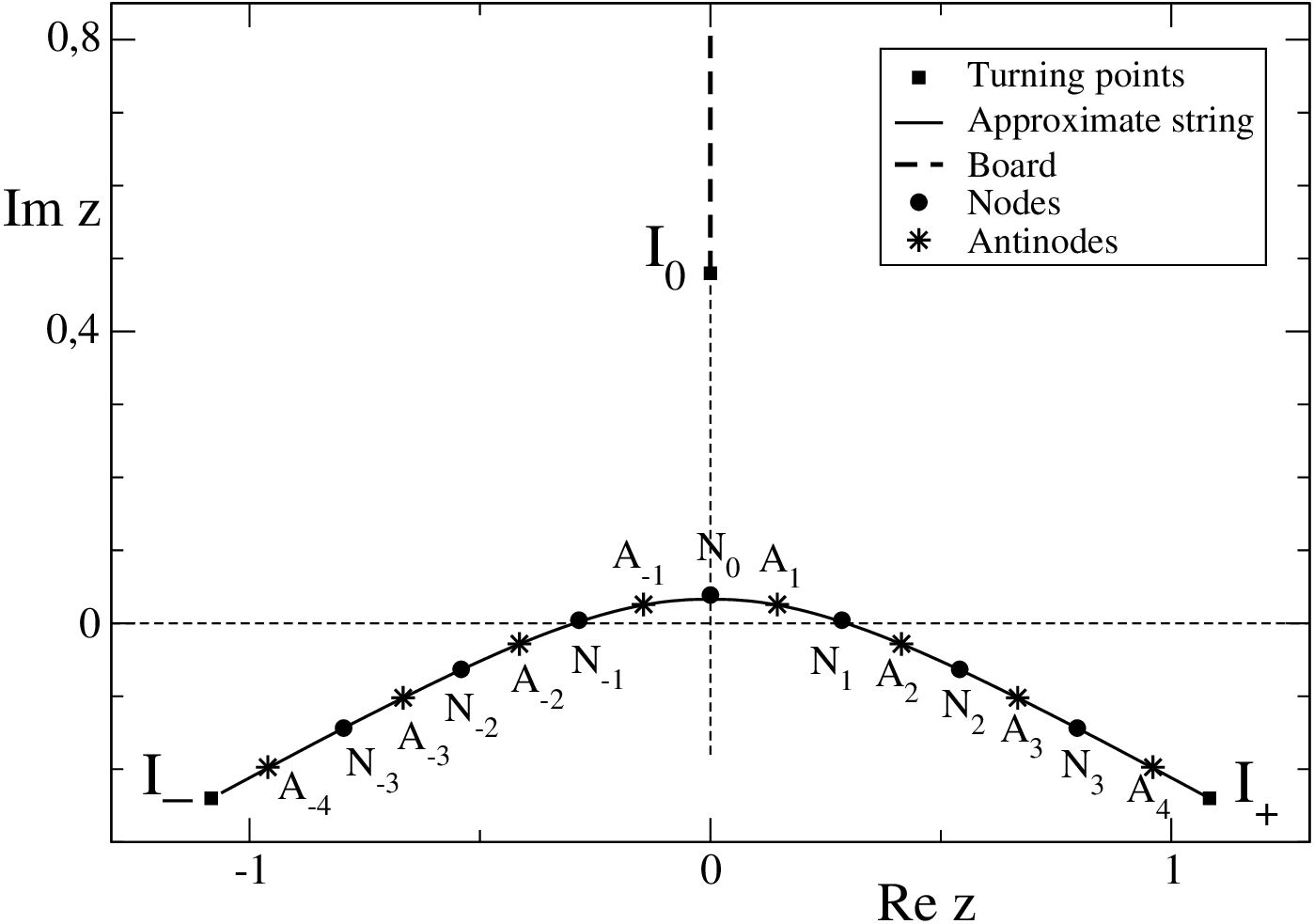}[h]}
	\caption{ $\hb>h_3^p$. The approximate monochord of $E_{7}(\hb)$  with the nodes and 
		antinodes.}
	\label{Nodi_Antinodi_059}	
\end{figure}

Following the process of crossing for decreasing  $\hb,$  just after the crossing we have the 
breaking of both the string and  the sequence of the nodes.
The limit of the critical energies, $E^c_{n,n}\ra E^c$ is an instability point of the Stokes 
complex.
 At the energy $E^c$ the exceptional Stokes line touches the short Stokes line  \cite{DT}, 
(Fig.1). 
We believe it is useful to try now to complete the picture of all the semiclassical  
behaviors of a level $E_m(\hb)$ in the complex plane. 
It is clear that for non-real $\hb$ other crossings of the same type are possible. Since the 
\textit{PT}-symmetry is lost, we admit that the indexes $j,k$ of the two levels undergoing crossing 
are different, and their sum  $j+k$ is not necessarily even. On the other side,  at least one of 
the nodes  of the state $\psi_{j+k+1}(\hb)$ must be unstable for the crossing of 
$E_{j+k+1}$ with a level $E_m$, $m< j+k+1.$ The simplest  possible 
generalization to the non-real $\hb$ case is obtained if we assume that 
exactly one of the nodes is unstable as in the symmetric case, so that 
the level $E_{j+k+1}(\hb)$ crosses the  level $E_{j+k}(\hb)$.
\bigskip

{\textsc{Conjecture C2$'\,$}} \textit{The standard sequences of the nodes   of the states 
$\psi_{j+k}(\hb),\,$ $\psi_{j+k+1}(\hb),\,$ for large $|\hb|$, with a suitable labeling, are 
respectively, 
$$S_{j+k}=(A_{-j-1}, 
~N_{-j},\,\,...\,\,,A_{-2},~N_{-1},~A_0~,N_1,~A_2,\,\,....\,\,,N_k,~A_{k+1}),$$
$$S_{j+k+1}=(A_{-j-1}, 
~N_{-j},\,\,...\,\,,~A_{-1},~N_0,~A_1,\,\,....\,\,,N_k,~A_{k+1}).$$ 
There exists a  parameter $\hb_{j,k}^a$ such that the 
antinode $A_0$ of the state $\psi_{j+k}(\hb_{j,k}^a)$ coincides with the turning point  
of the board
$I_0$.  There exists a  parameter $\hb_{j,k}^p$ such that the node $N_0$ of the state 
$\psi_{j+k+1}(\hb_{j,k}^p)$ coincides with the turning point of the board $I_0$. 
The sequence of the nodes of the state at the crossing is, 
$$S^c_{j,k}=(A_{-j-1}, 
~N_{-j},\,\,...\,\,,N_{-1},~A_{-1},~A_1,~N_1,\,\,....\,\,,N_k,~A_{k+1}),$$ 
and the sequences of the nodes of the states $\psi_j^-$,  $\psi_k^+$ are respectively 
$$S^-_j=(A_{-j-1},~N_{-j},\,\,...\,\,N_{-1}, 
~A_{-1}),\,\qquad\,S^+_k=(A_1,~N_1,\,\,...\,\,,N_k,~A_{k+1})\,.$$}
\begin{figure}
	\centerline{\includegraphics[height=200pt]{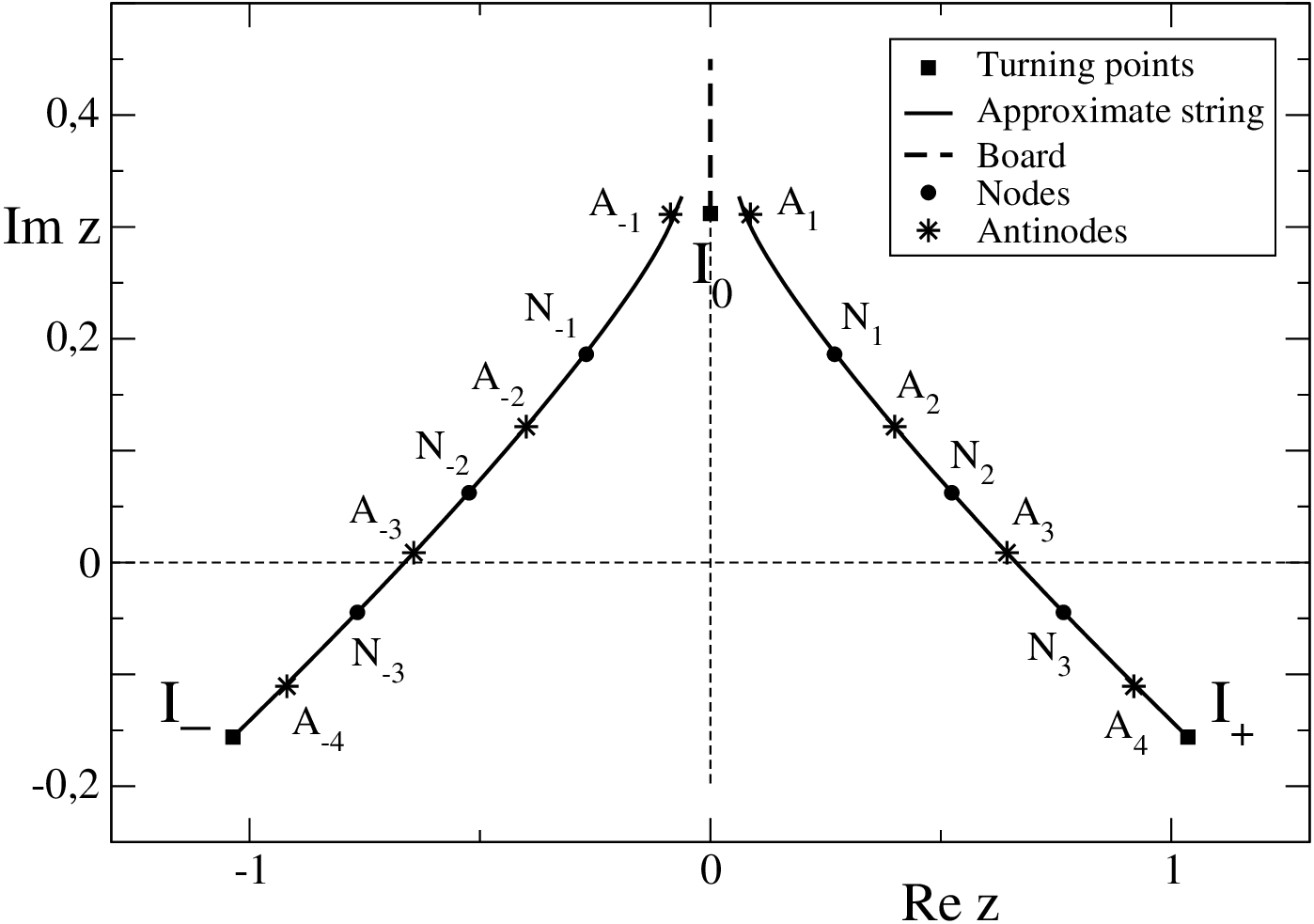}[h]}
	\caption{ $h_3<\hb<h_3^a$. The approximate monochord of $E_{6}(\hb)$ with the nodes and 
		antinodes.}
	\label{Splitting}	
\end{figure}
\begin{figure}
	\centerline{\includegraphics[height=200pt]{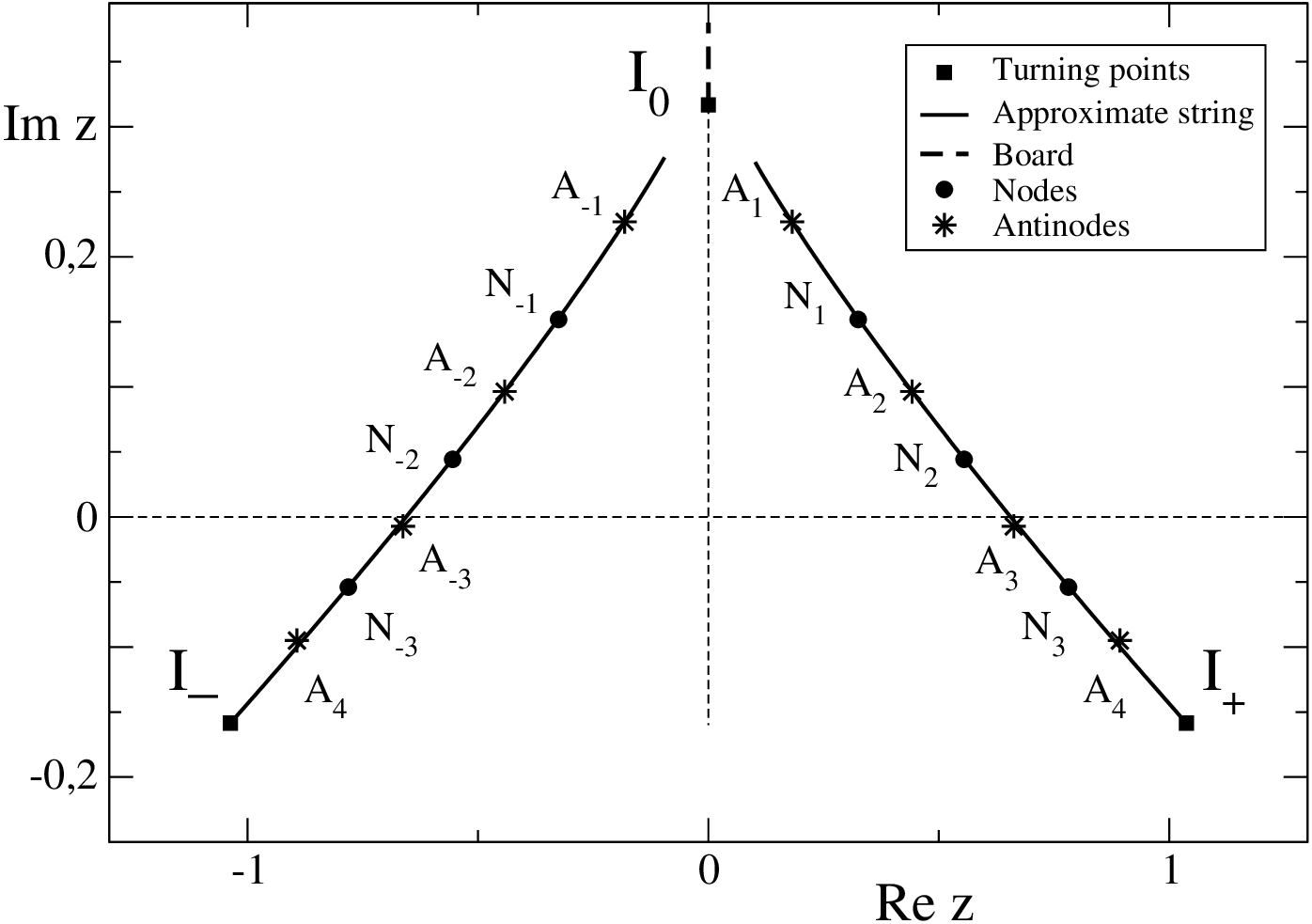}[h]}
	\caption{$h_3<\hb<h_3^p$. The approximate monochord of $E_{7}(\hb)$ with the nodes and 
		antinodes.}
	\label{Piccolo_h}	
\end{figure}

Again, if we follow the process of crossing for decreasing  $|\hb|,$ 
after the crossing  we have the breaking of the string and of the sequence of nodes and
the limit of the critical energies, $E^c_{n,\delta_n}\ra E^c(\delta)\,$  as 
$\,n=[(j+k)/2]\ra\infty\,$ 
and $\,\delta_n=(k-j)/n\ra\delta\,$, is an instability point of the Stokes complex.
Thus,   for   complex  parameter, the following crossings are possible:
the two  levels,
$E_{j+k+(1/2)\pm(1/2)}(\hb)$, $\,(j,k)\in\mathbb{N}^2\,$,  cross at   
$\hb_{j,k}\in\mathbb{C}^0$ giving two semiclassical levels  $E^-_{j}(\hb)$ and $E^+_{k}(\hb)$ 
for small $|\hb.|$

If we assume, according to the Hypothesis H2 (to be more precisely formulated in the 
following), that no crossing different from the above ones is possible  
and if we use Hypotheses H1  we obtain recursively the full picture of the Riemann sheets
of the levels (Theorem 3). 
The level $E_m(\hb),\,$  well defined and holomorphic for large $|\hb|$, 
has different behaviors for $\hb\ra 0$ along different paths
tangent to the real axis at $0$. Near the origin there exists a  partition 
of $\mathbb{C}^0$  into a finite number of stripes, ordered for increasing 
imaginary part,
$$ S_{-}^{m,0},~S_{m,0}^{m-1,0},~S_{m-1,0}^{m-1,1},~S_{m-1,1}^{m-2,1},\,\,....\,\,
S_{1,m-2}^{1,m-1},~S_{1,m-1}^{0,m-1}, ~S_{0,m-1}^{0,m},~S_{0,m}^+$$ 
where the behavior of $E_m(\hb)$
is  respectively  expressed by the following semiclassical levels in the same order 
$$E_m^-(\hb),~E_0^+(\hb),~E_{m-1}^-(\hb),~E_1^+(\hb),\,\,...\,\, 
E_1^-(\hb),~E_{m-1}^+(\hb),~E_0^-(\hb),~E_m^+(\hb).$$
\begin{figure}
	\centerline{\includegraphics[height=200pt]{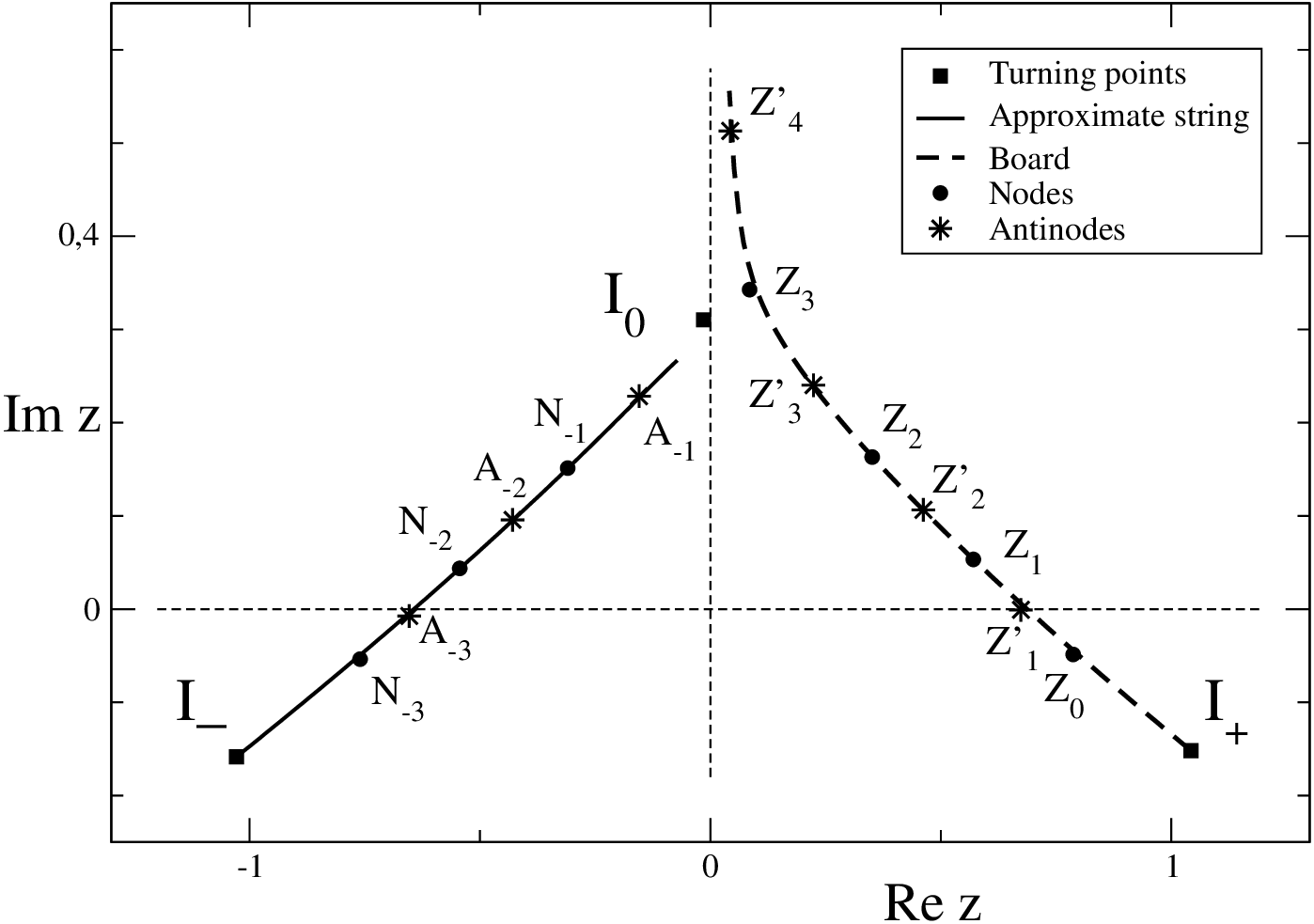}[h]}
	\caption{ $0<\hb<h_3$. The approximate monochord of $E_3^+(\hb)$ with the nodes, the 
		antinodes, the zeros and the stationary points.}
	\label{NodiAntinodiComplessi}	
\end{figure}%

To conclude, we give a brief summary of the paper content. In Sec. 2 we deal with the behavior of 
the levels and of the nodes for large values of $|\hb|$.
We prove a confinement of the nodes for large $\hb>0$, the positivity of the spectrum for large 
$\hb$, the reality of the states on the imaginary axis and we also consider the instability  
of the imaginary node of the odd states.
In Sec. 3 we study the behavior of levels and states in the semiclassical limit and we
show that for small $\hb>0$ the imaginary axis is free of zeros and the nodes are 
bounded. We prove then the total P-symmetry breaking at the crossing. In Sec. 4 we determine the 
possible quantization rules and we consider the Riemann surfaces of the levels in a neighborhood
of the real axis of $\hb$. The general crossing rule and level Riemann surfaces are considered
in Section 5. In the final Section 6 we introduce the string and the board, by which we
determine the sequences of nodes and antinodes. We add two  Appendixes concerning  
the semiclassical series expansions and some considerations on the numerical aspects.
\bigskip

 \section{Behavior of levels and nodes for large $|\hb|$}
\bigskip

The necessity of level crossing comes from the comparison of levels and states for large $\hb$ and 
small $\hb>0$. In this section we begin by investigating the principal features of levels and states 
for large $\hb.$
\bigskip
 
{\normalsize \emph{{\textbf{$($a$)$ Analyticity and confinement  of  the nodes for large $|\hb|$}}}}
\bigskip

In order to fix the number $m$ of nodes of a state $\psi_m(\hb)$ for large $|\hb|$, we prove a 
confinement of the nodes, so that the nodes are the only zeros in a certain region of the 
complex plane.
For $\hb>0$ large, it is convenient to use the representation  (\ref{KH}) and the Hamiltonian 
$K(\al)$  
so to have uniformly bounded  energy and nodes. Let us consider  the level 
$\hat{E}_m(0)$, $m\in\mathbb{N},$ of $K(0) \equiv K(\al=0)$, corresponding to the level 
$E_m(\hb)$ 
of $H_\hb$ at the limit of $\hb=+\infty$ according to (\ref{RHT}). 
It is important to observe that the scaling given in 
(\ref{RHT}) is regular,  with a positive (although unbounded) scale  $\lambda=\hb^{2/5}$ that 
maintains the phases on the complex plane. Due to (\ref{RHT}) the level $\hat{E}_m(\al)$ is 
positive for $\al\in\mathbb{R}$ and $|\al|$.  We prove now a confinement of the 
nodes and of the other zeros.
We translate the operator $K(0)$ by $x\ra x+iy,$ and we let
 $$ K_y(0)=p^2+i(x+iy)^3=
 p^2+i(x^2-3y^2)x+y^3-3yx^2=p^2+V_y(x)\,\,\, .$$
 We then apply the Loeffel-Martin method \cite{LM} to a level $E=\hat{E}_m(0)>0$, with a state
 $\psi=\hat\psi_m(0)$:
 \begin{eqnarray}
 &{}& -\Im \,[\overline{\psi}(x+iy)\pa_x\psi(x+iy)]=\int_{x}^\infty\Im 
 V_y(s)\,|\psi(s+iy)|^2ds=\cr
&{}&\phantom{xxx} \int_{x}^\infty(s^2-3y^2 )
s|\psi(s+iy)|^2ds=-\int^{x}_{-\infty}(s^2-3y^2 ) s|\psi(s+iy)|^2ds\neq 0
\nonumber
 \end{eqnarray}
 for $\,\pm x\geq \sqrt{3}\,|y|,\,$ $y\in\mathbb{R}\,$.
 In this case  we have a rigorous confinement of the region of   the nodes
 $$\mathbb{C}_\sigma=\{z=x+iy,\,\,y<0,\,|x|<-\sqrt{3}\,y\}\,\,
 \subset\,\,\mathbb{C}_-=\{z\in\mathbb{C},\,\Im 
 z<0\}.$$ 
The same confinement extends to all $\al>0$ and we see that the $m$ zeros of the 
 state $\widehat{\psi}_m(\al)$ on $\mathbb{C}_-$ are stable in the limit $\al\ra+\infty,\,$ 
 namely they are nodes by definition. 
 Previous computations of the nodes \cite{BBS}   suggest that  the present 
 confinement may be  sharp.
Since we know the analyticity of every level $\tilde E_m(\beta)$ of 
$H(\beta)$ as long as the $m$ nodes of the state $\tilde\psi_m(\beta)$ are in $\mathbb{C}_-$, 
we want to look what happens at $\beta=+\infty.$ In the case of the Hamiltonian 
$H(\beta)=p^2+x^2+i\sqrt\beta x^3,$ $\beta>0,$ by the scaling $x\ra\lambda x$
with positive $\lambda=\beta^{-1/10}$, we have 
$$H(\beta)\sim \beta^{1/5}(p^2+\beta^{-2/5}x^2+ix^3)$$ 
so that the level $\tilde E_m(\beta)$ of $H(\beta)$, has the behavior, \be\tilde E_m(\beta)\sim 
\beta^{1/5}\hat E_m(0)\,\,\, \textrm{as} \,\,\,\beta\ra+\infty.\label{AIB}\ee 
Thus, the operator $K(0)$
gives the asymptotic behavior of the spectrum of both the family of operators $H(\beta)$ for 
$\beta\ra+\infty$ and the family of operators $H_\hb$ for $\hb\ra+\infty$. As the nodes of the 
state $\hat\psi_m(0)$
are in $\mathbb{C}_-$,  the regularized levels $\beta^{-1/5}\tilde E_m(\beta)$ are real analytic up 
to 
$\beta=+\infty$ \cite{GM}. This means the absence of level crossings at $\al=0.$ But  
a level crossing of $\hat E_m(\al)$ is possible at a parameter $\al=\al(m)<0.$ 
Therefore the level $\hat E_m(\al)$ 
is real  analytic and the nodes of $\psi_m(\al)$ are in $\mathbb{C}_-$ for $\al$ in 
$[\al(m),+\infty)$.
At the same time, the level $E_m(\hb)$ is real  analytic and the nodes of $\psi_m(\hb)$ are in 
$\mathbb{C}_-$ for $\hb$ in $[\hb(m),+\infty)$, $\al(m)=-\hb(m)^{-4/5}.$
Thus, we state a result:
 \bigskip
 
 \textbf{Lemma 1} \textit{The level $E_m(\hb)$  is real  analytic and all the zeros of  
 $\psi_m(\hb)$  in $\mathbb{C}_\sigma$ as well as in  $\mathbb{C}_-$ are its m nodes for $\hb\in 
 [\,\hb(m),+\infty)$, 
 where $\hb(m)\geq 0$.
All the other infinite zeros are in,
 $$\mathbb{C}_B=\{z=x+iy,\,\,y>0,\, |x|<\sqrt{3}\,y\}
 \,\,\subset\,\,\mathbb{C}_+=\{z\in\mathbb{C},\,\Im 
 z>0\}.$$
Therefore the levels $\hat E_m(\al)$ are real analytic for $\al\in\mathbb{R}$ for $|\al|$ small.
\label{lemma1}}
\bigskip

We will see that actually such zeros in $\mathbb{C}_B$ are imaginary.   
\vfill\break

{\normalsize \emph{{\textbf{$($b$)$ Positivity of the levels  and reality of the states on the\\ 
\phantom{\textbf{$($b$)$}}			imaginary axis for large $\hb>0$ }}}}
\bigskip\bigskip

The  level  $\hat{E}_m(\al)$,  $m\in\mathbb{N}$ of $K(\al)$ is  analytic in a 
neighborhood  of the 
origin $U\subset \mathbb{C}$ \cite{GM,S}. Since it is real analytic   for $\al<0$, it is real 
analytic also 
in  $U\bigcup\mathbb{R}$ \cite{SH}.  The positivity of the real part of the levels comes 
from the numerical range and, in particular, from the kinetic energy
$$\Re \hat{E}_m(\al)=\Re 
\langle\widehat{\psi}_m(\al),K(\al)\widehat{\psi}_m(\al)\rangle
=\langle\widehat{\psi}_m(\al),p^2\widehat{\psi}_m(\al)\rangle\,>\,0,$$
where $\psi_m(\al)$ is the corresponding normalized state. Also the level $E_m(\hb)$ is 
real analytic and positive for $\hb>0$ large enough.
Thus, we have proved:
\bigskip

 \textbf{Lemma 2} \textit{Any  given level $E_m(\hb)$ is  positive for  large positive $\hb$.
 }\label{lemma2}
\bigskip
 We now extend the analysis of the analytic states on the complex plane. Let us consider 
 $y\in\mathbb{R}$ and the  translation  $ f(x)\ra f(x+iy)$, so that the \textit{PT}-symmetric 
 Hamiltonian    becomes the  {\textit{isospectral}} \textit{PT}-symmetric Hamiltonian 
 \be H_\hb(y)=h^2p^2+i(x^3-(3y^2+1)x)-(3yx^2-y^3-y)\sim H_\hb.\label{RKHK1}\ee
 The eigenfunction $\psi_{n,y}(x)=\psi_n(x+iy)$ with {{\it{real}} 
 eigenvalue $E_n$  can be  taken \textit{PT}-symmetric on the $\mathcal{H}_y$ representation},
 \be PT\psi_{n,y}(x)=\overline{\psi}_{n,y}(-x),\label{PX}\ee 
 so that, in particular, 
 $$\,\,\,\psi_{n,y}(0)
 =\overline{\psi}_{n,y}(0)=\psi(iy).$$
 Therefore, we have proved the following,
 \bigskip
 
 \textbf{Lemma 3}
 \textit{ If the level $\,E_m$, $m\in \mathbb{N},\,$ is positive then the state $\psi_m(z)$ 
 extended 
 as 
 an entire function on the complex plane, is $P_xT$-symmetric, 
 \be (P_xT\psi_m)(x+iy)=\overline{\psi}_m(-x+iy)={\psi}_m(x+iy),\,\,\quad\forall 
 x,y\in\mathbb{R},\label{PX1}\ee and the set of  its zeros  is $P_x$-symmetric. In particular,  for 
 a choice of the gauge, the state is real on the imaginary axis,  \be \Im\psi_m(iy)=0,\,\,\forall 
 y\in\mathbb{R}.\label{RSI1}\ee }\label{lemma3}
\bigskip

{\normalsize \emph{{\textbf{$($c$)$ The nodal analysis of the process of crossing}}}}
\bigskip

Let  us to fix $\hb>0$ large enough and let  $E=E_m(\hb)$  be a positive level  of the Hamiltonian 
(\ref{RKHKC1}) with a corresponding state $\psi_m(z)$. Now, by the complex dilation $z\ra iz$, we 
consider  the  Hamiltonian on the imaginary axis:
 \be   H^r_\hb=-\hb^2\frac{d^2}{dy^2}+\tilde{V}(y)\,\sim\,-H_\hb,\,\,\qquad
 \tilde{V}(y)=-y^3-y,\label{MHB}\ee  
 {well defined} by the $L^2$ condition on the $x$-axis, here 
 playing the role of the imaginary axis. The Hamiltonian $H^r_\hb$  has the same  spectrum as 
 $-H_\hb,$  so that  $-E=-E_m(\hb)<0$ is one of its eigenvalues (Lemma 2). The corresponding state  
 $\phi_m(y)=\psi_m(iy)$  can be taken  real for $y$ real. In particular, for $y>0$ 
 large, because of the two fundamental solutions and the reality property, we can write 
 \be\phi_m(y)\sim 
 \frac{C}{\sqrt{p_0(E,y)}}\cos (p_0(E,y)+2\pi\al),\,\,\label{AAPI0}\ee 
 with a $C>0$ and where
 $$p_0(E,y)=\sqrt{y^3+y-E},\qquad \al\in\mathbb{R}/\mathbb{Z}\,.$$ 
 For $\,-y>0\,$ large, we have a real  combinations  of the two fundamental solutions, 
 \be\phi_m(y)\sim \frac{C'}{\sqrt{p_0(E,y)}}\,(\exp (p_0(E,y))+a\exp 
 (-p_0(E,y))\,),\,\,\label{AAPI1}\ee 
with a $\,C'>0,\,$ $p_0(E,y)=\sqrt{-y^3-y+E},\,$  $a\in\mathbb{R}.$
  We consider together the two states $\psi_m(z)$,    $[m/2]=n\in\mathbb{N}$,  for a fixed $\hb\geq 
  \hbar_n$. Both the states have  $n$ nodes on both the half-planes $\mathbb{C}^\pm$ and 
  are 
  distinguished by the number of imaginary nodes  for $\hb>0$ large.
The whole process of crossing for $\hb\geq \hbar_n$ can be studied by the behaviors of the states 
$\psi_m(z)$ with energy $E=E_m>0$,  on the imaginary semi-axis, called the continuation of 
the board,\be B^c(E):=\{z=iy,\, -\infty<y<\tilde{y}(E)\},\label{CBOIA}\ee where the imaginary 
turning point is $I_0=i\,\tilde{y}(E)$. This means that we consider each one of the two formal states 
$\phi(y):=\phi_m(y)$, of the representation (\ref{MHB}),  for 
$\,y\leq 
{\tilde{y}}(E).\,$
  For $\hb>0$ large, we have two possible behaviors of the state $\phi(y)$ of 
(\ref{MHB}). Let us recall that for $y$ in a open interval of the semi-axis 
$-\infty<y<\tilde{y}(E)$, if  a state $\phi(y)$ is positive then it is  convex;
 if it is negative then it is  concave. On the other side,  for $\,y>\tilde{y}(E)\,,$ where an 
 eigenfunction $\phi(y)$ is positive  it is also concave
and where it is negative it  is also convex.  Since we can consider $\phi(y)$
positive decreasing for $y \ll\tilde{y}(E),$ there are only two cases: 

\hspace{12.0pt} $\phantom{i}i$) \hspace{1.0pt} the existence of one  zero  on $B^c(E)$,  

\hspace{12.0pt} $ii$) \hspace{1.0pt} the absence   of  zeros on $B^c(E)$. 

Let us remark that  
$\tilde{y}(E)>0$  so that, by Lemma 1, a possible  node on the imaginary axis should be in 
$B^c(E)$ for large $\hb>0$.
A state without imaginary nodes can have one or zero antinodes.
We can state the result:
\bigskip 

\textbf{Lemma 4} \textit{Let $\,E(\hb)$, with $\hb>0\,$, be a positive level with a 
corresponding \textit{PT}-symmetric state $\psi(\hb)$. When $\hb$  is large 
enough the state $\psi(\hb)$ can have one or zero imaginary nodes.}
\bigskip
 
For the existence of an imaginary node for large $\hb>0$ we consider a state with labeling 
$m=2n+1$, continued to all  $\hb>\hb_n>0$. 
An imaginary node is indeed unstable since it can cross the turning point $I_0$ at a parameter 
$\hb_n^p>\hb_n.$ For large $\hb,$ a state $\psi(\hb)$  with an imaginary node  has no imaginary 
antinodes and a a state $\psi(\hb)$  without imaginary nodes can have one or zero 
imaginary antinodes. It is possible and actually necessary that at a parameter $\hb_n^a>\hb_n$ the 
imaginary antinode disappears and two non imaginary antinodes of $\psi_{2n}(\hb)$ are generated. 
We clarify this fact by a simplified example. Let 
$\psi_{2n}(z)=i({z^3}/{3}-\es z)+c$, $c\neq 0$. We have $\psi'(z)=i(z^2-\es)=0$ at the stationary 
points $z_\pm=\pm\sqrt{\es}$ and $\psi''(z)=2iz=0$ at $z=0=I_0$.  
For $\es=\hb_n^a-\hb>0$  there are two non-imaginary antinodes and 
for $\es<0$ there the only antinode $-i\sqrt{|\es|}$ in lower complex half-plane.
\bigskip

\section{Semiclassical limit, confinement of the nodes and the crossing rule}
\bigskip

We now study the behavior of the levels and the states in the semiclassical limit. From the 
comparison with the large $\hb>0$ behavior we will prove the necessity of the level crossings.
\bigskip

{\normalsize \emph{{\textbf{$($a$)$ From semiclassical to perturbation theory and semiclassical\\ 
\phantom{\textbf{$($a$)$}} limit  of the nodes}}}}
\bigskip

Let $\hb\in\mathbb{C}^0$ with $|\hb|$ small.  Some transformations are necessary 
in order to use the results of \cite{GM}
for the localized states. We consider Hamiltonian $H_\hb$ with two wells at $x_\pm=\pm1/\sqrt{3}$.
We make the unitary translations centering on one of the wells or the other one, $x=x_\pm+y=y\pm1/\sqrt{3},$ getting the new Hamiltonians:
 $$H_\hb^\pm=\hb^2p^2+i(y^3\pm\sqrt{3} y^2)\mp \textsl{E},\,\qquad\textsl{E}= i 
 \frac{2}{3\sqrt{3}}.$$
We make the suitable dilations in order to use the  perturbation theory \cite{GM}. We put 
\be
y=\lambda^\pm(\hb) z,\,\qquad\lambda^\pm(\hb)=\exp(\mp i\pi/8)3^{-1/8}\sqrt \hb\,
\label{dila}\ee
and we get
\be H_\hb^\pm\sim \hb c^\pm H(\beta^\pm(\hb))\mp \textsl{E}\phantom{xxxxxxxx}\label{PSR1}\ee
where
$$ c^\pm=3^{1/4}\sqrt{\pm i},\,\,\qquad  \beta^\pm(\hb)=\exp(\mp i5\pi/4) 3^{-5/4}\hb,$$
and where $H(\beta)$ is
\be H(\beta)=p^2+x^2+i\sqrt{\beta}x^2.\phantom{xxxi}\label{HB2}\ee 
Let us notice that the parameters 
$\beta^\pm(\hb)$ are not in the cut plane $$\mathbb{C}_c=\{z\in\mathbb{C};\,\,z\neq 0,\,\,|\arg 
z|<\pi\},$$ so that 
we cannot use all the results of \cite{GM}; nevertheless we can use some of the results of 
\cite {C}. It is 
clear from the perturbation theory that we have the semiclassical behavior of the levels,
\be E_n^\pm(\hb)=\mp i \textsl{E}+ \hb c^\pm (2n+1)+O(\hb^{2}),\,\,\hb>0.\label{0B}\ee
In the  perturbation theory of Hamiltonian (\ref{HB2}) we have the relevant fact that  in the 
semiclassical limit all the nodes of the state $\tilde{\psi}_n(\beta^\pm(\hb))$ go to the short 
Stokes line $\,[\,-\sqrt{2n+1},\sqrt{2n+1}\,\,].$
In the semiclassical limit this corresponds respectively to the wells $x_\pm$ of the nodes of 
the semiclassical states $\psi_n^\pm(\hb)$, 
\bigskip

\textbf{Lemma 5} \textit{Both the states 
$\psi_n^\pm(\hb)$ have $n$ zeros tending to the points  $x_\pm$, respectively, as 
$\hb\ra 
0^+$}.
\bigskip

We will prove a stable confinement  of the zeros of both  the states $\psi_n^\pm(\hb)$ 
in $\mathbb{C}^\pm$ (\ref{CPM}) respectively, so that such zeros coincide with the nodes. Thus, 
we will prove that no crossing between the levels of the same set $\{E_n^-(\hb\}_{n\in\mathbb{N}}$ 
or of the 
same set  $\{E_n^+(\hb\}_{n\in\mathbb{N}}$ can occur, contrary to crossings of the levels of 
$\{E_n^-(\hb)\}_{n\in\mathbb{N}}$ 
with the 
levels of $\{E_n^+(\hb\}_{n\in\mathbb{N}}$ that are indeed possible. 
\bigskip

{\normalsize \emph{{\textbf{$($b$)$ The confinements of the nodes and the crossing rule }}}}
\bigskip

 Let us consider  $\hb>0$ and a level $E\in\mathbb{C},$ with the corresponding  state $\psi(z),$ 
 and $\psi(iy)=\phi(y),$ $z,y\in \mathbb{C}$.
 We transform the Hamiltonian (\ref{MHB}) by imaginary translations: 
 $$ H^r_\hb(x)=-\hb^2\frac{d^2}{dy^2}+\tilde{V}(y-ix),\,$$
 \begin{eqnarray}
 &{}&\tilde{V}(y-ix)=\Re \tilde{V}(y-ix)+i\Im \tilde{V}(y-ix)=-(y-ix)^3-(y-ix)\spazio{0.8}\cr
 &{}&\phantom{\tilde{V}(y-ix)}=-y^3+3x^2y-y+i(x(3y^2+1)-x^3)
 \nonumber
 \end{eqnarray}
   where $\,\Im \tilde{V}(y-ix)=(x(3y^2+1)-x^3\,)$ with level $\,-E\,$, for a fixed $x\neq 0$. We 
   consider a state, $$\phi_x(y)=\phi(y-ix),\,\,\,\,\,n\in\mathbb{N},$$
  with the well known asymptotic behavior(\ref{AAPI0}), 
  $$\phi_x(y)\sim 
  \frac{C}{\sqrt{p_0(E,w)}}\cos (p_0(E,w)+\theta),\,\quad w=y-ix,\,\quad y\ra+\infty,$$ 
  for a $\,C>0,\,$  $\theta\in\mathbb{R}/2\pi\mathbb{Z}\,$ and
  $$|\phi_x(y)|^2=O(|y|^{-3/2})\, \,\quad\textrm{for}\,\quad y\ra+\infty.$$
Since the dominant term  is  bounded and real, we have,	$$\Im (\bar{\phi}(y)\pa_y\phi (y))\ra 0 
,\,\quad \textrm{as} \,\quad y\ra +\infty.$$
	We consider the Loeffel-Martin formula    in order to generalize to our problem the expression of the imaginary part of a shape resonance:
 \be\,\Im \,(\hb^2\overline{\phi}(y)\pa_y\phi(y))
 = -\Im E\,\int^{\infty}_{y}
 |\phi(s)|^2ds,\,\,\forall y\in \mathbb{R},\label{AN1}\ee
 where the integral in (\ref{AN1}) exists and is bounded for the semiclassical behavior. Thus we 
 state 
 the result:
\bigskip

 \textbf{Lemma 6} \textit{Let us consider the non-real levels $E^\pm_n(\hb)$ at a fixed value of 
 the parameter $\hb<\hbar_n$. The corresponding states  $\psi_n^\pm(z)$ are different 
 from $0$  on the imaginary axis and, being   entire functions,  they are free of zeros 
 in a  neighborhood of the imaginary axis. Obviously, the width of this neighborhood is not 
 uniform  at infinity. }
\bigskip

 We next apply the Loeffel-Martin method \cite{LM} generalized to the case of diverging integrals:
\begin{eqnarray}
&{}&\Im \,[\hb^2\overline{\phi}_x(y)\,\pa_y\phi_x(y)]=\Im 
\,[\hb^2\overline{\phi}_x(y_0)\pa_y\phi_x(y_0)]+\spazio{0.8}\cr
&{}&\phantom{\Im \,[\hb^2\overline{\phi}_x(y)]}\int_{y_0}^y(x(3s^2+1-x^2)+\Im E)
|\phi_x(s)|^2ds\ra+\infty,
\label{AN}
\end{eqnarray} 
as $y\ra+\infty$ for fixed $y_0,x\in\mathbb{R},$ $x\neq 0.$ We know that   the zeros, for  
large $|z|$, 
have the asymptotic direction $\arg z\ra \pi/2$ \cite{GM}. Let  $E\in\mathbb{C}_\pm$ be a non 
real level with state $\psi(z)$ of the Hamiltonian $H_\hb$ for a fixed $\hb>0$. In the regions 
$$\Omega^\pm=\{z=x+iy\in \mathbb{C}^\pm,\,\,x^2\leq 3y^2+1,\,\,y>0\,\},$$ 
for $E\in\mathbb{C}_\pm$ respectively, there are no zeros for large $y$. 
 Let for instance $\Im E>0$,  $y_0>0$; by (\ref{AN}) we have the absence of a zero at $(x,y)$ for  
 $0<x<1$ and $y>y(x)$ for a function $y(x)>y_0$.
The function  $y(x)$ is not uniformly bounded for $x$  small, but this is not a problem  because of Lemma 6.
This means that the large zeros are on $\mathbb{C}^\mp$ if $E\in\mathbb{C}_{\mp}$, respectively.  
In the limit of $\hb\ra \hb_n^-$ the energies $E_n^\pm(\hb)$ become positive, and the large zeros 
of 
$\psi_n^{\pm}(\hb)$ become  imaginary.
We are then able to state a stronger condition on the asymptotics of the zeros:
\bigskip

 \textbf{Lemma 7} \textit{Since $E_n^\pm(\hb)\in\mathbb{C}_\mp$,  the $n$ nodes of the two 
 states  $\psi_n^\pm(\hb)$, near $x_\pm$ for small $\hb$, stay respectively   
 in $\mathbb{C}^\pm$  for all $\hb<\hb_n$. 
 Since the state $\psi_n^+(\hb)$ $(\psi_n^-(\hb))$ is the only one to have $n$ nodes in 
 $\mathbb{C}^+$ $(\mathbb{C}^-)$, 
 the two functions $E_n^\pm(\hb)$ are analytic for $0<\hb<\hb_n.$
 At the crossing limit, the two levels $E_n^\pm(\hb)$ with the states $\psi_n^\pm(\hb)$, coincide. 
 The state  $\psi_{n,n}^c$ at the crossing is $PT$-symmetric and has $2n$ non-imaginary 
 zeros conventionally considered the only nodes. The large zeros are imaginary.}
\smallskip

\textbf{Proof\phantom{..}} We have, $\psi_n^+(\hb)=PT\psi_n^-(\hb)$ for $\hb<\hbar_n$, and 
$\psi_n^\pm(\hb)\ra\psi_{n,n}^c$ as $\hb\ra \hbar_n^-$, so that $\psi_{n,n}^c=PT\psi_{n,n}^c.$ The 
state $\psi_n^+(\hb)$ has only $n$ zeros in $\mathbb{C}^+$ and the state $\psi_n^-(\hb)$ has only 
$n$ zeros 
in $\mathbb{C}^-$. Since at the limit $\hb\ra \hbar_n^-$ these zeros cannot diverge or become 
imaginary,  all the limits of the non-imaginary  zeros of both the state $\psi^\pm(\hb)$ are all the non-imaginary zeros 
of the limit state $\psi_{n,n}^c.$ 
\smallskip

We say conventionally that the $2n$ non-imaginary zeros are the  nodes of  $\psi_{n,n}^c.$  
\bigskip

Consider  a state $\psi_m(\hb)$, for $\hb>\hbar_n$ having limit $\psi_{n,n}^c$ as $\hb\ra 
\hbar_n^+$. We have:
\bigskip

 \textbf{Lemma 8} \textit{Let $\psi_m(\hb)$, for $\hb>\hbar_n$, be a generic state having limit 
 $\psi_{n,n}^c$ as $\hb\ra \hbar_n^+.\,$ For $\hb>\hb_n,\,$  $\psi_m(\hb)$  has 
 exactly  $2n$ non-imaginary zeros, possible nodes, stable at $\hb_n$.  Taking into account
 the possible 
 existence of one imaginary node, the   number $m$ of its  nodes is not greater than  $2n+1$.}
\medskip 

\textbf{Proof\phantom{..}} For $\hb>\hb_n$, both the states $\psi_m(\hb)$ are \textit{PT}-symmetric 
and the 
corresponding levels $E_m(\hb)$ are positive (Lemma 2 and 3). Because of the symmetry  and the 
simplicity of the spectrum, a non-imaginary zero cannot become imaginary and an imaginary zero    
cannot leave the imaginary axis.  Due to (\ref{AN}),  a non-imaginary  zero of 
$\psi_m(\hb)$, with energy $E=E_m(\hb),$ can go to  infinity along a path asymptotic to the 
imaginary axis at infinity.   But at any fixed 
$\hb>\hb_n$ the state 
$\psi_m(\hb)$  has the following behavior in a neighborhood of  the 
imaginary axis (\ref{AAPI0}), 
$$\psi_m(z)=\phi_m(w)\sim \frac{C}{\sqrt{p_0(E,w)}}\cos 
(p_0(E,w)+\al),\,\,w=y-ix,\,\, y\ra+\infty,$$ 
for a $C>0,$ $p_0(E,w)=\sqrt{w^3+w-E},$  
$\al\in\mathbb{R}/\mathbb{Z},$ so that it is free of zeros  for a  small $|x|\neq 0,$ 
$x\in\mathbb{R}$ and  $y>0$ large enough.\\ 
The number of  non-imaginary nodes of both the states $\psi_m(\hb)$ is $2n$ as the state 
$\psi^c_{n,n}$. All the non-imaginary zeros can go to the half plane $\mathbb{C}_-$ for large 
$\hb>0$, as the nodes do. 
We know that only one of the imaginary zeros can be a node (Lemma 4). Thus, the maximum number of 
nodes is $2n+1,$ whereas the minimum number is $0.$
Because of  the independence of the two states $\psi_m(\hb)$ having limit $\psi_{n,n}^c$ as $\hb\ra 
\hbar_n^+,$ the maximal values of the pair of numbers $m$ is $(2n,2n+1)$.  
\bigskip
\begin{figure}[t]
	\centerline{\includegraphics[height=180pt]{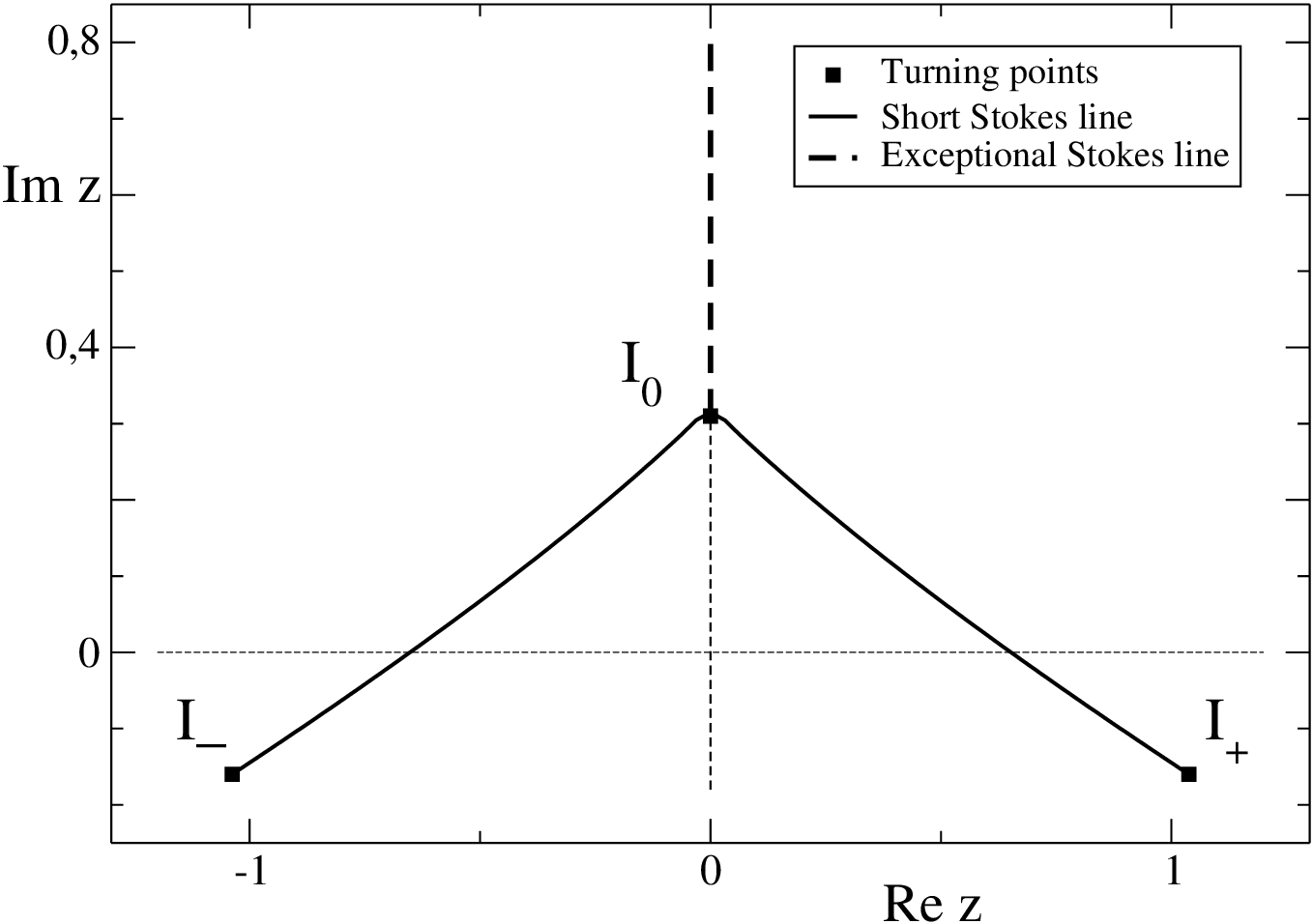}}
	\caption{$\hb=0$. The monochord, the subset of the Stokes complex, consisting of the short 
		Stokes line (string) and the exceptional Stokes line (board), at the critical energy 
		$E^c=0,35226..$.}
	\label{StokesCritico}
\end{figure}

Actually,  considering the sequence of pairs of levels $E_m(\hb)>0$ obtained by the crossings for 
large $\hb>0$, only the the maximum values of the pairs of number $m$, $(2n,2n+1)$ are compatible 
with the uniqueness of each level. Only the sequence of pairs, 
$$(E_0,E_1),(E_2,E_3),(E_4,E_5),...,$$
gives exactly the full sequence of levels $E_0,E_1,E_2,...$. The  imaginary node of the state 
$\psi_{2n+1}(\hb)$ 
is always imaginary but can well coincide with the lowest imaginary zero of $\psi_{2n}(\hb)$ at 
$\hb=\hb_n.$

  Thus, we state the following,
\bigskip
\begin{figure}
	\centerline{\includegraphics[height=180pt]{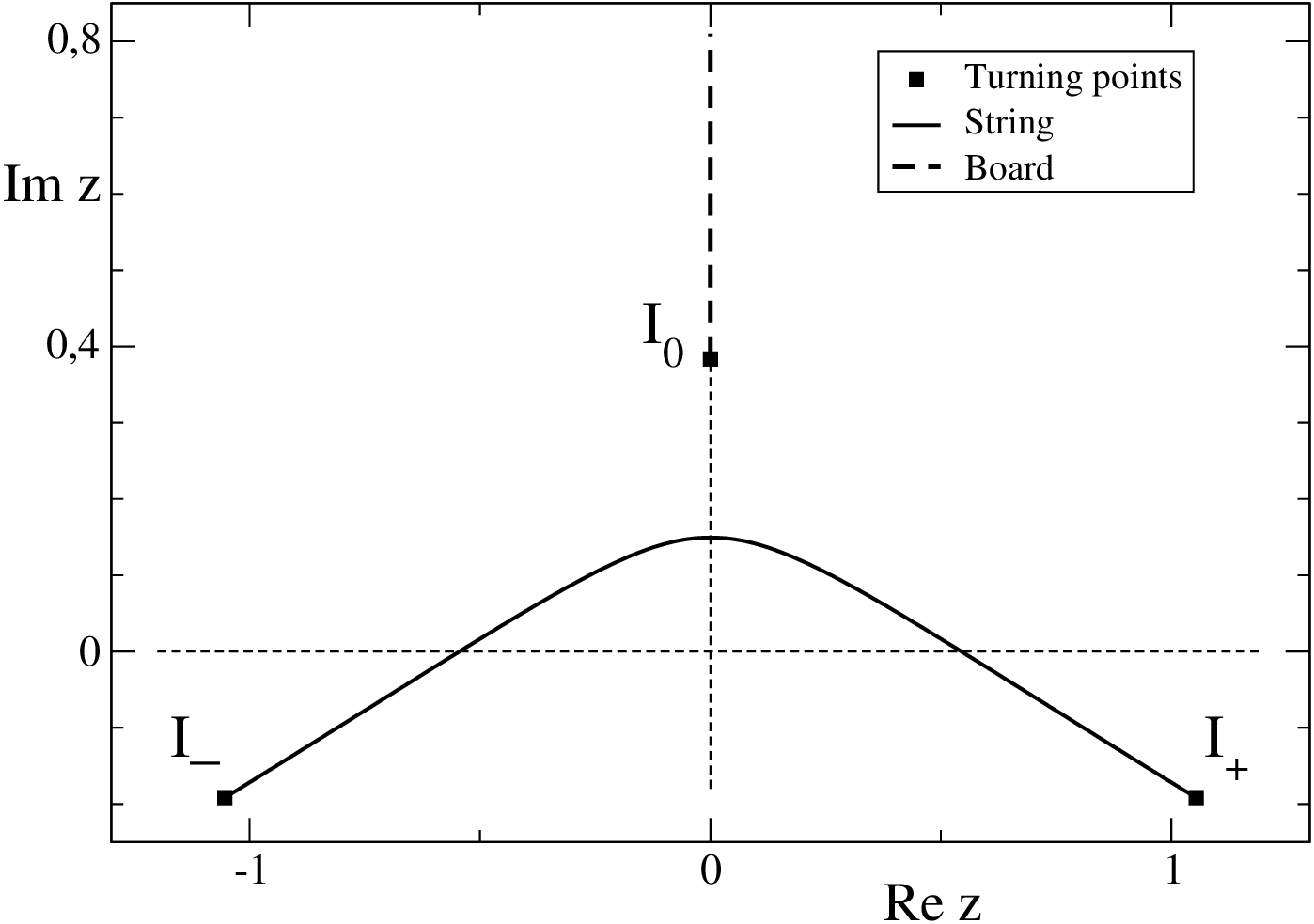}}
	\caption{$\hb$=0. Instability of the monochord for a positive variation of the energy: 
		$E>E^c$.}
	\label{StokesRe}	
\end{figure}

\textbf{Theorem  1} \textit{For each $n\in\mathbb{N}$, there exists a parameter $\hbar_n>0$ and 
a crossing at $\hbar_n$. The two levels $E_{2n+(1/2)\pm(1/2)}(\hb)$ separated for 
$\hb>\hbar_n$, and 
the two levels $ E_n^\pm(\hbar_n)$ separated for $\hb<\hbar_n$, crosses at $\hbar_n>0.$
The two states $\psi_{2n+(1/2)\pm(1/2)}(\hb)$, $\hb\gg \hbar_n,$ have a set of $2n$ 
non-imaginary nodes. }
\medskip
    
\textbf{Proof\phantom{..}} The existence of the crossings is necessary because of the 
positivity of the analytic functions $E_m(\hb)$ for large $\hb>0,$ and the non reality of the 
analytic functions $E_n^\pm(\hb)$ for small $\hb>0$. In particular, if seen from 
$\hb\leq \hbar_n$, we have a crossing between the levels $E_n^\pm(\hb)$ when they becomes real
and  equal. The crossing between the levels $E_{2n}(\hb)$, $E_{2n+1}(\hb)$ is possible because the 
stability of the $2n$ non-imaginary nodes of both the states $\psi_{2n+(1/2)\pm(1/2)}(\hb)$ and 
the instability of 
the imaginary node of the state $\psi_{2n+1}(\hb)$. Because of the $P_xT$-symmetry of both the 
states $\psi_{2n+(1/2)\pm(1/2)}(\hb)$, they have $n$ nodes in both the half-planes 
$\mathbb{C}^\pm$. The continuation to $\hb<\hb_n$ of the $n$ nodes 
in $\mathbb{C}^\pm$ are the nodes of the states $\psi_n^\pm(\hb)$ in $\mathbb{C}^\pm$, 
respectively. 
\bigskip
		
\textsc{{Remark 1}} \textit{ The zeros on the upper half-plane for large $\hb>0$ are all 
imaginary.}
\bigskip

This statement strengthens  the confinement of the zeros for large $\hb>0$ obtained above. 
It ensues from the 
result that all the non-imaginary zeros are nodes, and all the nodes are in the lower 
half-pane for large $\hb>0.  $
\bigskip

{\textsc{Conjecture C3}} \textit{The sequence $\hbar_n$ has a vanishing limit for $n\ra\infty$}.
\bigskip

This conjecture is based on the semiclassical and the exact semiclassical theory. 
It is related to the conjecture that $ n \hb_n$ and $2 n \hb_n$ tend to the 
action integral of single well $J^+(E^c,0)=J^-(E^c,0)$ and of double well 
$J_2(E^c,0)$, respectively, as $n\ra\infty$ (\ref{CC}), (\ref{CC2}). The instability of the nodes is related to the
contact of the string with the board and the instability of the string  at $E=E^c$, 
$\hb=0.$ For the possibility of proving  this conjecture by perturbation theory 
see \cite {CCG,N,ZJ}.

Also the behavior of the isolation distance for large $\hb$ and large $n$
agrees with this conjecture.
\medskip
\begin{figure}[h]
	\centerline{\includegraphics[height=180pt]{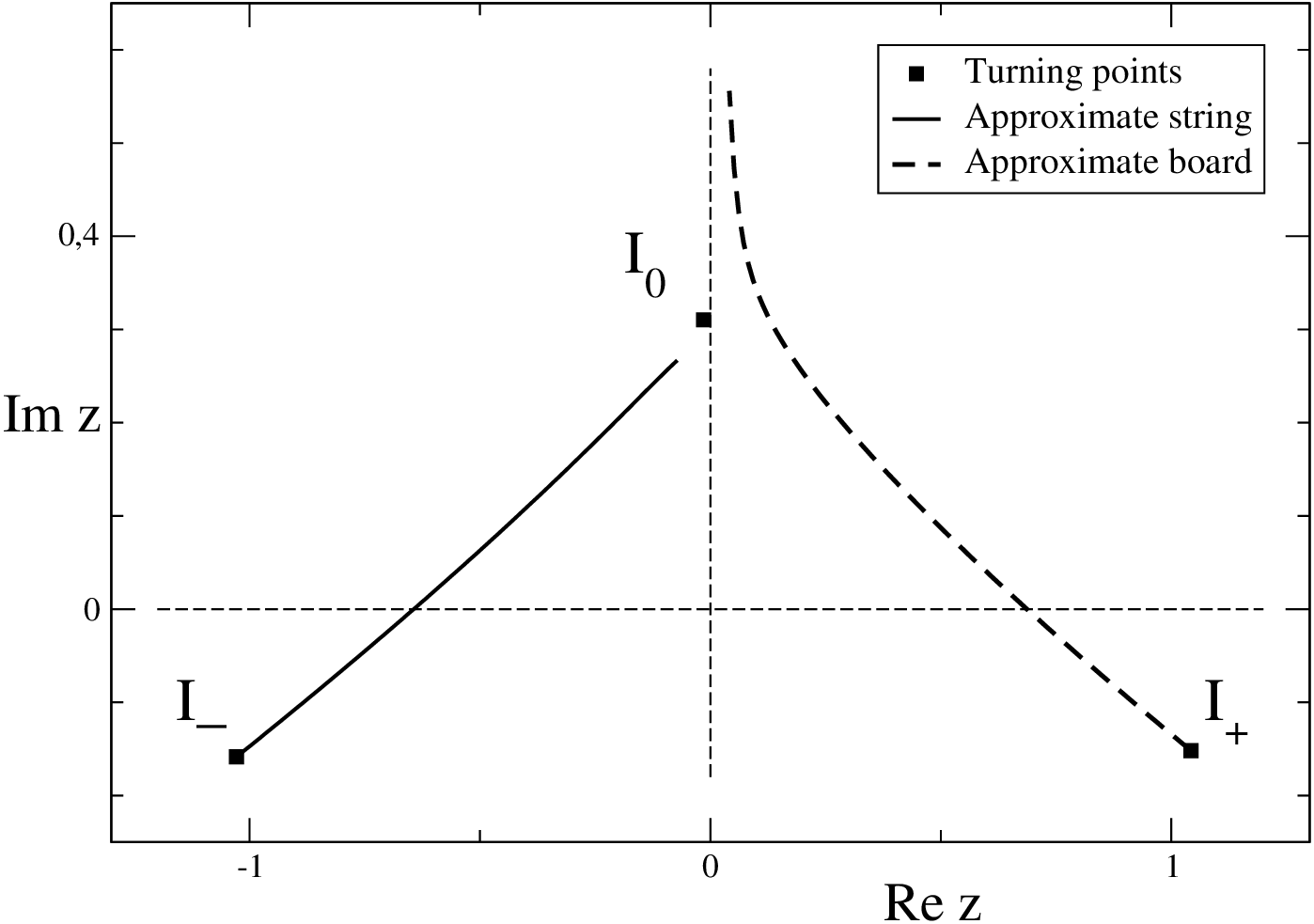}}
	\caption{$\hb=0$. Instability of the monochord for a complex variation of the energy: $E\neq 
		\bar{E}$.}
	\label{StokesIm}	
\end{figure}

{\normalsize \emph{{\textbf{$($c$)$ The total P-asymmetry at the crossing }}}}
\bigskip

 We have seen that the states $\tilde{\psi}_n=\tilde{\psi}_n(0)$ of $H(\beta)$ at fixed $\beta=0,$ 
 have definite parity: $P\tilde{\psi}_n=(-1)^n\tilde{\psi}_n$. This means that $|\tilde{\psi}_n|^2$ 
 is $P$-symmetric, and the expectation value of the parity is 
 $\,\langle\tilde{\psi_n},P\tilde{\psi_n}\rangle=(-1)^n$. We want to prove that the state 
 at the  crossing, $\psi_n^c=\psi_n(\hbar_n)$, $n\in\mathbb{N},$  has vanishing mean value of
 the parity,
 $\langle\psi_n^c,P\psi_n^c\rangle=0,\,$ so that it is totally $P$-asymmetric in the sense 
 that $\psi_n^c$  is orthogonal to $P\psi_n^c$.

 We have a crossing of $\,E_n^\pm(\hb)\,$ at $\,\hb=\hbar_n\,$ when $\,\Im E_n^\pm(\hb)=0.\,$ For 
 $0<\hb<\hbar_n,$ the two clamped points of $\psi_n^\pm$ are $(I_\mp,I_0)$ respectively.
 At the crossing, we have $P_x$ symmetry of the turning points, so that $I_-=\bar{I_+},$ 
 $I_0=-\bar{I_0}$.\\
 Let   $\,H=H_\hb,\,$  $\,H_\hb^*=\bar H=H_{\bar{\hb}},\,$ with two levels \,$E_j=\bar E_j\,$ and 
 states 
 $\,\psi_j,\,$ $j=1,2\,$. Then 
 $$H\psi_1=E_1\psi_1,\,\,\,\qquad \bar H\bar\psi_2=E_2\bar\psi_2,$$ 
 so that 
 \be\langle\bar\psi_2,H\psi_1\rangle=E_1\,\langle\bar\psi_2,\psi_1\rangle=
 E_2\,\langle\bar\psi_2,\psi_1\rangle\label{PE}\ee
 and, by subtraction 
 $$0=(E_2-E_1)\,\langle\psi_1,\bar\psi_2\rangle\,.$$ 
 Let now to vary the 
 semiclassical parameter $\hb$, so that:
 $$ 0=(E_2(\hb)-E_1(\hb))\,\langle\psi_1(\hb),\bar\psi_2(\hb)\rangle,$$ 
 for $\hb>0.$ If $E_1(\hb)\neq E_2(\hb)$ 
 for $\hb> \hbar_n,$  and $E_1(h^+_n)=E_2(h^+_n)=E,$ $\psi_1(h^+_n)=\psi_2(h^+_n)=\psi,$ we have 
 \be 
 0=\langle\psi,\bar\psi\rangle=\langle\psi,P\psi\rangle=\int_{\mathbb{R} }\psi^2(x)dx.\label{CA}\ee
 We have thus proved:
 \bigskip
\begin{figure}
	\centerline{\includegraphics[height=180pt]{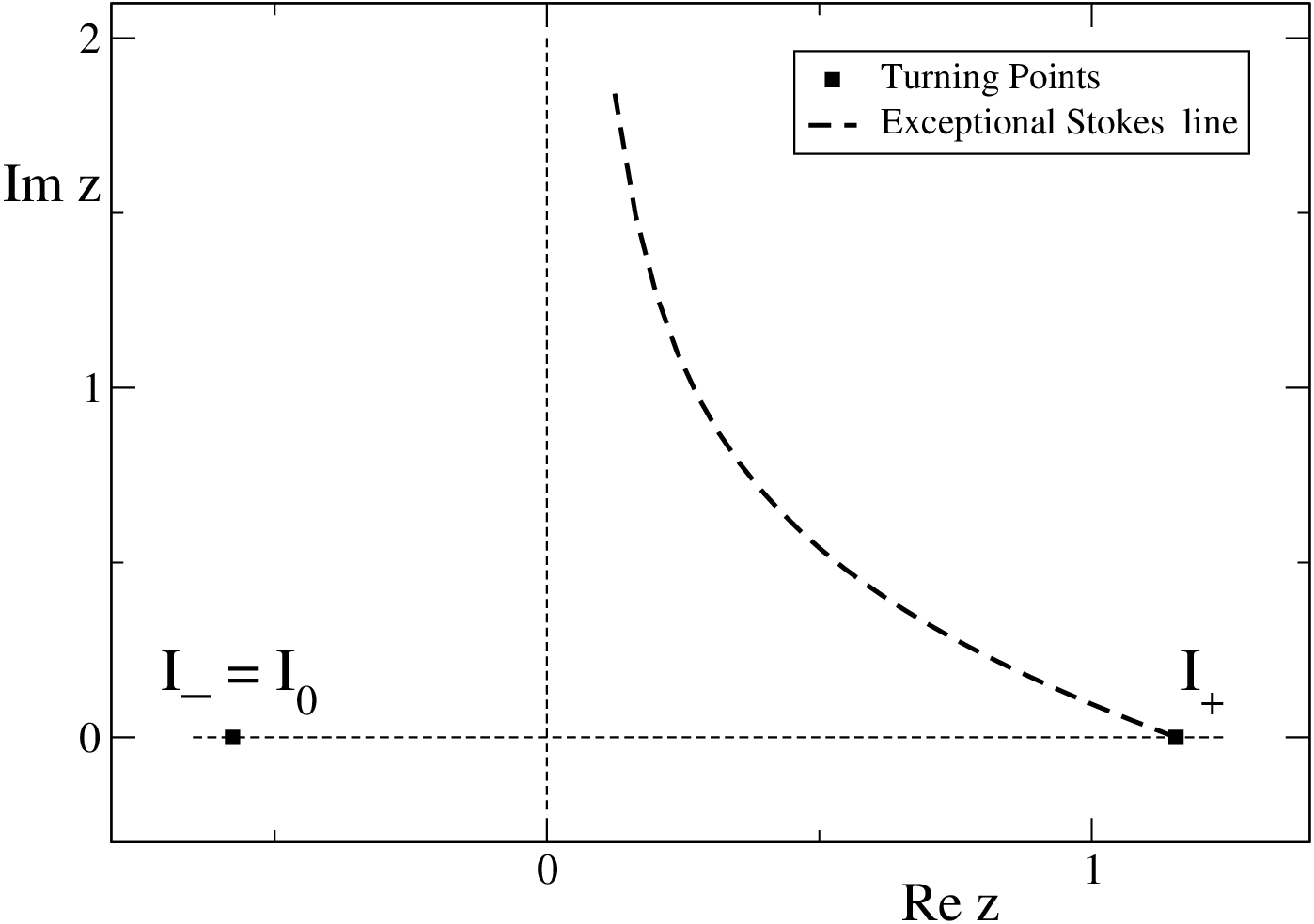}[h]}
	\caption{$\hb=0$. The monochord at the semiclassical energy $E_n^-(0)=i2/3\sqrt{3}$, $\forall 
		n\in\mathbb{N}$. The string is the point $I_-=I_0=x_-.$}
	\label{PuntoDoppio}	
\end{figure}

 \textbf{Lemma 9}\textit{
 	The \textit{PT}-symmetric state at the crossing point, 
 	$${\psi}^c_{n,n}=\psi_{2n+1}(h^+_{n})=\psi_{n,n}(h^+_{n})=PT{\psi}^c_{2n},$$
 	is completely P-asymmetric, namely  
 	$\langle{\psi}^c_{n,n},P{\psi}^c_{n,n}\rangle=0.$}
 \bigskip
 
Considering the states as eigenfunctions of the Hamiltonian $K(\alpha)$, the
state $\psi=\hat\psi_{2n+1}$ is odd in the sense that it gives a negative mean value of the parity 
operator , $\langle\psi,P\psi\rangle<0$ tending to $-1$ in the limit $\alpha\ra +\infty.$
Conversely, the state $\psi=\hat\psi_{2n}$	 is even since $\langle\psi,P\psi\rangle>0$ and 
tends to $1$ in the limit $\alpha\ra +\infty\,.$ Thus, the crossing of two levels with states of  
opposite parities for large $\al$, is possible.
\bigskip

\section{The  boundedness of the levels and the Riemann surfaces}
\bigskip

In this section  we examine the possible types of
quantization rules excluding the divergence of the levels. We then consider the properties of the Riemann surfaces of the eigenvalues in 
the neighborhood of the real axis.
\bigskip

{\normalsize \emph{{\textbf{$($a$)$ The quantization rules and the boundedness of the levels}}}}
\bigskip

There are two types of rigorous quantization rules for $\hb>0$ giving the boundedness of the levels 
for bounded $\hb.$ Moreover, there is another rigorous quantization rule for large $\hb>0.$
We have seen that there are two kinds of confinement of the nodes depending on two conditions for 
the energy: if the energy level satisfies the condition $E\in\mathbb{C}^\mp$ then the set of $n$ 
nodes  of the corresponding state $\psi$ is confined on $\mathbb{C}^\pm$
 respectively. Thus, both the levels $E_n^\pm(\hb)$, $\hb<\hb_n,$ satisfy 
 the  unique conditions on the phase and on the nodes, do not cross and are analytic. 
We have two kinds of quantization rules for a fixed  $\hb< \hbar_n$,
 giving the levels $E_n^\pm$ and the states $\psi_n^\pm$. At $\hb=\hbar_n$, the two 
 levels become positive and we have the crossing. 
 
 Suppose there exist two continuations of both 
 levels $E_n^\pm(\hb)$, $n\in\mathbb{N},$ from $\hb<\hbar_n$ to $\hb>\hbar_n.$ For the moment we 
 maintain  the 
 same names $E_n^\pm(\hb)$ for the continuations of the levels, even if  such  continuations should 
 be distinguished by different labeling
We know that both continuations of the energy levels are positive and both  continuations of the 
states 
have $n$ nodes in both $\mathbb{C}^\pm.$
There exist  two regular circuits $\gamma^\pm$ such that
$$ P_x\gamma^+=\gamma^-,\qquad\gamma^\pm=\partial \Omega^\pm ,$$
where $\Omega^\pm$ are regular regions large enough, with 
$$\Omega^\pm\subset\mathbb{C}^\pm=\{x+iy,\, \pm x>0,\, y\in\mathbb{R}\}\,.$$
and the exact quantization conditions read 
$$
\frac{1}{2i\pi}\oint_{\gamma^\pm}\frac{\psi'(z)}{\psi(z)}dz= n\,. 
$$
We can better write 
\be J^\pm(E,\hb):=
\frac{\hb}{2i\pi}\oint_{\gamma^\pm}\frac{\psi'(z)}{\psi(z)}dz+\frac{\hb}{2}=\hb \Bigl( 
n+\frac{1}{2}\Bigr), 
\label{CC}\ee
if $\psi(z)=\psi_n^\pm(\hb,z)$ and  
$E=E_n^\pm(\hb)\in\mathbb{C}^\mp\,$ 
respectively. 

In particular, for small $\hb>0,$ and fixed  $n\in\mathbb{N},$ the quantization rules 
(\ref{CC}) become the semiclassical quantization conditions for $E=E_n^\pm(\hb)$ , 
\be J(E,\hb)=
\frac{1}{2i\pi}\oint_{\gamma^\pm}p_0(E,z)dz+O(\hb^2)=\hb \Bigl( n+\frac{1}{2}\Bigr), \label{CC1}\ee
where 
$p_0(E,z)=\sqrt{V(z)-E},$  and the paths $\gamma^\pm$ shrink around the short Stokes 
line.
Both the quantization conditions (\ref{CC}) at $\hb= \hb_n$ give the  same  solution 
$E_n^c$, $\psi_n^c$, and for $\hb>\hb_n$ both  give the both the solutions $E_m(\hb)$, 
$\psi_m(\hb)$, 
$[m/2]=n\in\mathbb{N}.$

We distinguish the two solutions by the selecting condition 
\be  E_{2n+1}(\hb)>E_{2n}(\hb).\label{C1}\ee 
Therefore both functions $E_m(\hb)$ are analytic for $\hb>\hb_n.$
For a fixed, large  $\hb\gg\hb_n$ we have the exact  quantization rules,
\be J_2(E,\hb) :=
\frac{\hb}{2i\pi}\oint_{\Gamma}\frac{\psi'(z)}{\psi(z)}dz+\frac{\hb}{2}=\hb ( m+\frac{1}{2}), 
\label{CC2}\ee
where the solutions are, 
$$\psi(z) =\psi_m(\hb,z),\,\quad E=E_m(\hb),\,\quad m=2n\,\,\textrm{or} 
\,\,2n+1,\,\quad \Gamma=\Gamma_m=\partial \Omega_m,$$ 
and where $\Omega_m\subset \mathbb{C}_-$ is large enough in order to contain all the $m$ nodes.

These 
quantization conditions (\ref{CC}), (\ref{CC2})  yield the boundedness and the continuity
of the levels even at the crossing point $\hbar_n$.
\bigskip

{\textbf{Lemma 10}} \textit{$(a)$ The two functions   $E_n^\pm(\hb),$ are analytic   for 
$\hb<\hbar_n$. The two functions   $E_m(\hb)$, $[m/2]=n$ are 
 analytic for $\hb>\hbar_n$.\\ $(b)$ Let $E(\hb)$,  be one of the two functions 
 $E_n^\pm(\hb),$  
 $n\in\mathbb{N}$  for $\hb\leq \hbar_n$  with one of its two continuations   $E_m(\hb)$, 
 $[m/2]=n$ 
 for 
 $\hb>\hbar_n$. The function $E(\hb)$ is bounded and continuous on $\mathbb{R}_+$ and is analytic 
 with 
 a square 
 root singularity at ${\hbar}_n$.}
\medskip

 \textbf{Proof\phantom{..}} The point ($a$) is proved by the exact quantization conditions 
 (\ref{CC}) 
 with the selection condition (\ref{C1}) for $\hb>\hbar_n$.

 We prove by absurd point($b$). We assume the divergence of $E(\hb)$ at $h^c\gg\hbar_n$ where the 
 $m$ 
 nodes of the corresponding state are in $\mathbb{C}_-$. The extension to the general case is 
 simple. 
We consider the operator, $$\frac{H_\hb-E(\hb)}{|E(\hb)|}\sim {\hat \hb}^2\,p^2+ix^3-i\delta 
x-\eta,$$ 
by a scaling $x\ra\lambda x,$ $\lambda=|E|^{1/3}$, where ${\hat\hb}=\hb |E^{-5/3}|,$ 
$\delta=|E|^{-2/3}$, $\eta=E/|E|$, $|\eta|=1$.  For small ${\hat\hb}>0,$ by  simply putting 
$\delta=0,$
we have the semiclassical quantization condition, 
\be
\frac{1}{2i\pi}\oint_{\Gamma_m}\sqrt{iz^3-\eta}\,\,dz={\hat\hb}\Bigl(m+\frac{1}{2}\Bigr)
+O({\hat\hb}^2),
\label{CC3}\ee where $\Gamma_m=\pa\Omega_m$ and all the $m$ nodes are in $\Omega_m\subset C_-.$ It 
is easy to see that (\ref{CC3}) can be satisfied only if
$\eta\ra 0$ as ${\hat\hb}\ra0.$\\

{\normalsize \emph{{\textbf{$($b$)$ The Riemann surfaces  near the real axis }}}}
\bigskip

Let us consider the sector  (\ref{SIH}) on the $\hb$ complex plane,
$$\mathbb{C}^0=\{\hb\in \mathbb{C};\,\hb\neq 0\,, \arg(\hb)<\pi/4\},$$
 and the Riemann sheet 
$\mathbb{C}^0_m$ of the level $E_m(\hb)$, $n=[m/2]$, defined  in $\mathbb{C}^0,$
 with a square root singularity at $\hbar_n$
 and a cut, $\gamma_{n,n}=(0,\hbar_n\,]\,$. We prove the following:
\bigskip
 
\textbf{Theorem 2}\textit{
The levels $(E_{2n+1}(\hb),E_{2n}(\hb))$ are analytic functions defined on the 
Riemann sheets $(\mathbb{C}^0_{2n},\,\,\mathbb{C}^0_{2n+1})$ respectively, both of them having  
only the cut  
$\gamma_{n,n}=(0,\hbar_n\,]$ on the real axis. The
positive analytic functions $(E_{2n+1}(\hb),E_{2n}(\hb)),$   with $E_{2n+1}(\hb)>E_{2n}(\hb)$ on 
$(\hbar_n,+\infty)$ take the following values at the boundaries of the cut: 
\be E_{2n}(\hb\pm i0^+)={E}^\pm_{n}(\hb),\,\,E_{2n+1}(\hb\pm i0^+)={E}^\mp_{n}(\hb),\,\,\forall\,\, 
0<\hb<\hbar_n.\label{SON}\ee
}
\medskip

\textbf{Proof\phantom{..}} Since both the functions $(E_{2n+}(\hb),E_{2n}(\hb))$ have a square root 
singularity at $\hb_n,$ and $$E_{2n+1}(\hb_n+\es)-E_{2n}(\hb_n+\es)=O(\sqrt{\es})>0,$$ for $\es>0$ 
small, $$\pm\Im(E_{2n+1}(\hb_n+\exp(\pm i\pi)\es)-E_{2n}(\hb_n+\exp(\pm i\pi)\es))<0,$$ and $\mp\Im 
E_n^\pm (h)>0,$ for $h<\hbar_n,$ we necessarily  have, $$E_{2n+1}(\hb_n+\exp(\pm 
i\pi)\es)={E}^\mp_{n}(\hb_n-\es),$$ $$E_{2n}(\hb_n+\exp(\pm i\pi)\es)={E}^\pm_{n}(\hb_n-\es).$$
\bigskip

\textsc{Remark 2} {Let us consider the crossing process along a path starting from $\hb=0$, 
turning around the singularity at $\hbar_n$
and going back at $\hb=0.$ The state $\psi_n^+(\hb)$ concentrated at $x_+$ at the beginning  of the 
path, becomes the state $\psi_n^-(\hb)$  concentrated at $x_+.$
Now, we consider a path starting at large $\hb>0,$ turning around $\hbar_n$ and going back to a 
large $\hb>0$.
An odd state $\psi_{2n+1}(\hb)$ at the beginning of the trip becomes an even state 
$\psi_{2n}(\hb)$: the imaginary node of the lower half plane  becomes an imaginary zero of the 
upper half plane.}
\bigskip

 \textsc{Remark 3} {The Riemann sheet $\mathbb{C}^0_0$ of the fundamental level has only 
 one cut $\gamma_{0,0}=[0,h_0]$ on $\mathbb{R}$ {\rm{\cite{DT}}}, and the discontinuity on the 
 cut is 
 defined by  
 the rule, \be E_{0}(\hb\pm i0^+)={E}^\pm_{0}(\hb),\,\,\forall\,\,\hb,\,\, 
 0<\hb<h_0.\label{S0N}\ee} 

 We recall, for instance, that $E_0^+(\hb)=E_0(\hb+i0)$, is defined as the limit from above for 
 small $\hb>0.$ This definition extends directly to all $\hb>0$ in the absence of complex 
 singularities. Formula (\ref{S0N})  means that the absence of other singularities involving the 
 function $E_0(\hb)$ is possible. Thus, by using the Hypothesis H1, we assume that in 
 $\mathbb{C}_0^0$ there is only the cut $\gamma_{0,0}$.
\bigskip

 \section{The general crossing rule for complex $\,{\boldsymbol{\hb}}\,$ and the  Riemann surfaces 
 	of  the levels}
\bigskip
 
 We extend the study to the general case of $\hb\in\mathbb{C}^0$ where it is more difficult to 
 prove the selection rules. The potential $V(x)=i(x^3-x)$ is \textit{PT}-symmetric with two wells 
 at $x_\pm=\pm 1/\sqrt{3}$ respectively. Let us consider the parameter $\al$ along a line defined 
 by a fixed $c\in\mathbb{R}$,  $\{ \al=-r+ic,\,\forall r>0\}$. The corresponding line on the  $\hb$ 
 complex  plane, 
 \be\gamma_c=\{\hb=(r+ic)^{-4/5}, \, r>0\},\label{LC}\ee 
 is tangent to the real axis at the origin.The choice of these kind of paths is arbitrary but is 
 justified by the  semiclassical analysis of the crossings in the $\al$ complex plane.
 
  For $\hb$ in the line $\gamma_c$, $c\neq 0$, we still have a  double well but  the 
  \textit{PT}-symmetry
  of the Hamiltonian  is broken. For a fixed $c\neq 0$ and a large $r>0$ we expect the existence 
  of the levels $E_j^\pm(\hb)$ and the $j$-nodes states $\psi_j^\pm(\hb)$ localized in 
  the $x_\pm$ well respectively \cite{C}.  Even if there are no crossings in $\gamma_c$ for a 
  $c\neq 0$, with $|c|$ small, it is clear that continuing the level $E_j^\pm(\hb)$ and the state 
  $\psi_j^\pm(\hb)$,  up to  $r>0$ small enough,  the state becomes delocalized and should change 
  name. The delocalized states for $|\hb|$ large are called  $\psi_{m}(\hb)$ for an $m\geq j$ to be 
  specified. In this case, we expect  that the continuation of $\psi_j^\pm(\hb)$ is $\psi_m(\hb)$ 
  with an $m\geq j$ to be discussed. On the other side, if we start with a $\psi_m(\hb)$, 
  $\hb\in\gamma_c$ for a  small $r>0,$ if $c>0$ is large enough,
  we expect to have no crossings for all $r>0$, and $\psi_m(\hb)$ becomes $\psi^+_m(\hb)$ for 
  $r>0$ large.

We now consider the Riemann sheet of the level $E_m(\hb)$ for large $|\hb|$ and continued on all 
the sector $\hb\in\mathbb{C}^0$. 
    We always assume the  minimality condition on the number of crossings (Hypothesis H1).
	 Thus we  extend the result of Lemma 9 end we assume,
	\bigskip 
 
 {\textsc{Hypothesis H2}}
   \textit{ The generalization of the crossings to non positive $\hb$ and different indexes  $ 
   (j,k)\in\mathbb{N}^2$ is the natural one. On one side  
   $E_{j+k}(\hb)$ crosses $E_{j+k+1}(\hb)$ and on the other side 
$E^+_{j}(\hb)$ crosses $E^-_{k}(\hb)$ at the same parameter
      $\hb_{j,k}\in\mathbb{C}^0.$} 
\bigskip    
    
    The Hypothesis H2, difficult to prove, is  the simplest 
      generalization of Theorem 1  concerning the case of $j=k=n$ for positive 
      $\hb_{n,n}=\hbar_n>0.$   We will see (Theorem 3) that with this rule we can  have a minimal 
      structure of singularities (in agreement with Hypothesis H1).   
	\\We define the Riemann sheet of the eigenvalue $E_m(\hb)$, $m\in\mathbb{N},$ holomorphic for 
	large $|\hb|$, with the minimal number of branch points and cuts for small $|\hb|$.\\ We call 
	$\gamma_\pm$ the boundary lines of the sector $\mathbb{C}^0,\,$ $\gamma_\pm=\sqrt{\pm 
	i}\mathbb{R}_+.$ 
	Because of the results of \cite{GM}, we have the identity of two definitions at each side: 
	$E_m(\hb)=E_m^\pm(\hb)$ for $\hb\in\gamma_\pm$ respectively. 

	Let us consider the Riemann 
	sheet $\mathbb{C}^0_0$ of $E_0(\hb),$ with  only one positive singularity at $\hb_{0,0}=\hb_0$ 
	as proven before (Theorem 2).  The cut  on the positive interval $\gamma_{0,0}=(0,\hb_0]$ 
	separates the behaviors of $E_0(\hb)$ defined by $E^\pm_0(\hb)$ as $\hb\ra 0$ in sectors 
	$S_{0,0}^{+},\, $ $S^{-}_{0,0},\,$ that is for $\,\pm\Im \hb>0$, respectively.
	
	The sheet $\mathbb{C}^0_1$ of $E_1(\hb)$ has the same positive singularity at $h_0$, with the 
	following behavior on the boundaries of the cut $\gamma_{0,0}=(0,h_0]$ (Theorem 2): 
	\be 
	E_{1}(\hb+i0^+)=E_0^-(\hb),\,
   \,\,E_{1}(\hb-i0^+)=E_0^+(\hb),\,\,\forall\,\,\,\,\,
   \hb\in\gamma_{0,0}=(0,h_0].
   \label{S1}\ee 
   In order to have the correct behavior as $\hb\ra 0$ at the boundaries of  the sector, 
   $\gamma^\pm$,
 it is necessary the existence of the other pairs of complex conjugated singularities 
 $\hb_{1,0},\hb_{0,1}$ with cuts on suitable arcs of lines, $\gamma_{1,0},\gamma_{0,1}$,  of the 
 type (\ref{LC}) from the origin to $\hb_ {1,0},\hb_{0,1}$ respectively,  so that we get the full 
 sequence of singularities 
  $$\hb_{1,0},~~\hb_{0,0},~~\hb_{0,1},$$
 ordered by increasing imaginary part, 
 and the corresponding cuts, 
 $$\gamma_{1,0},~~\gamma_{0,0},~~\gamma_{0,1}.$$ 
 The behavior of the 
 function $E_1(\hb)$ as $\hb\ra 0$ on the stripe $\,\,S^{0,0}_{1,0}\,\,$ between $\gamma_{1,0}$ and 
 $\gamma_{0,0}$ is given by the function called $E_0^+(\hb)$. 
 The behavior of the function 
 $E_1(\hb)$ as $\hb\ra 0$ on the stripe $\,\,S^{0,1}_{0,0}\,\,$ between $\gamma_{0,0}$ and 
 $\gamma_{0,1}$ 
 is given by the function called $E_0^-(\hb)$. The behavior of the function $E_1(\hb)$ as $\hb\ra 
 0$ on the stripe $\,\,S_{0,1}^+\,\,$ between $\gamma_{0,1}$ and $\gamma_{+}$ is given by the 
 function 
 called $E_1^+(\hb)$.  The behavior of the function $E_1(\hb)$ as $\hb\ra 0$ on the stripe 
 $\,\,S_-^{0,1}\,\,$ between $\gamma_{-}$  and $\gamma_{1,0}$ is given by the function 
 $E_1^-(\hb)$. In 
 particular
   \be E_{1}(\hb\pm i0^+)=E_0^\mp(\hb)\quad\forall\,\,\hb\in\gamma_{1,0}\,;
   \qquad E_{1}(\hb\pm i0^+)=E_1^\pm(\hb)\quad
   \forall\,\,\hb\in\gamma_{0,1}.\label{GCRS1}\ee 
   Thus, the possible crossings defined 
   by  the parameters $\hb_{0,1}$, $\hb_{1,0}$ (Hypothesis H2) are necessary and sufficient in 
   order to have the simplest Riemann sheet of $E_1(\hb)$. We see that Hypothesis H1 justifies 
   Hypothesis H2 in the sense that this is absolutely the simplest Riemann sheet of 
   $E_1(\hb)$.

   The sheet $\mathbb{C}^0_2$ of $E_2(\hb)$ is given by adding  the singularities 
   $\hb_{2,0},\hb_{0,2},$ and  substituting $\hb_{0,0}$ with  $\hb_{1,1}$ because of the Theorem 2, 
   so that we get the sequence of singularities on $\mathbb{C}^0_2$, 
   $$\hb_{2,0},~\hb_{1,0},~\hb_{1,1},~\hb_{0,1},~\hb_{0,2}.$$ 
   We see that the crossings defined by the 
   parameters $\hb_{2,0},\hb_{0,2},$ are necessary and sufficient for the self consistency of the 
   sheet $E_2(\hb)$.
   
 In the case of  the sheet $\mathbb{C}^0_3$ of $E_3(\hb),$ we still have the singularities 
 $\hb_{2,0},\hb_{1,1},\hb_{0,2}$, but  the singularities $\hb_{1,0},\hb_{0,1},$ are substituted by 
 the singularities $\hb_{2,1},\hb_{1,2}$ respectively. This substitution is necessary because of 
 the rule (\ref{GCRS1}), and in order to have the  definite  behaviors $E_1^\pm(\hb)$ in the 
 stripes 
 $\,\,S_{2,1}^{1,1}\,\,$ and  $\,\,S_{1,1}^{1,2}\,\,$ respectively. Moreover, we have to add the 
 new singularities 
 $\hb_{3,0},\hb_{0,3}$ so that we get the the sequence of singularities on $\mathbb{C}^0_3$, 
 $$\hb_{3,0},~\hb_{2,0},~\hb_{2,1},~\hb_{1,1},~\hb_{1,2},~\hb_{0,2}~~,\hb_{0,3}.$$
 Thus, the singularities at the values $\hb_{2,1},\hb_{1,2}$, $\hb_{3,0},\hb_{0,3}$ are all 
 necessary, and together with the previous ones, sufficient for a self consistent sheet of 
 $E_3(\hb)$. Hence all the crossing corresponding to the values 
 $\hb_{j,k}$ with $0\leq j+k\leq 3$ are necessary and sufficient. 
 
 Going on, we get the general sequence.
 The sheet $\mathbb{C}^0_j$ of $E_m(\hb)$, $m>3,$ has the expected  sequence of singularities
  $\{\hb_{j,k}\}_{j,k}$ ordered by the increasing values of the imaginary part,  
  \be 
  \hb_{m,0},~\hb_{m-1,0},~\hb_{m-1,1},~\hb_{m-2,1},\,....,\,\hb_{1,m-2},~\hb_{1,m-1},
  ~\hb_{0,m-1},~\hb_{0,m},\label{SOS}\ee
  where each one of the two indexes  follows the rules of  
  decreasing of an unity the first index or increasing of an unity the second index
 alternatively, starting from the first index. We expect that the crossing parameters $\hb_{j,k}$ 
 with $j+k=m$ are almost aligned along a vertical line, as the $\hb_{j,k}$ with $j+k=m-1$ are 
 almost 
 aligned along another vertical line. Actually, the parameters $\hb_{j,k}$ with $j+k=m-1$ are near 
 the parameter $\hb^p_{j,k}$ with $j+k+1=m$ where a node of $\psi_{m}$ coincides with a turning 
 point $I_0$. Hence all the crossing corresponding to the 
 parameters $\hb_{j,k}$ with $0\leq j+k\leq m\,$ are necessary and sufficient. 
\bigskip

\textbf{Theorem 3} \textit{By the  Hypotheses $\mathrm{H}1$, $\mathrm{H}2 $, we get the full 
picture of the 
Riemann 
sheets. Let us consider  the function  $E_m(\hb),\,$  holomorphic for 
$|\hb|$ large. The sector $\mathbb{C}^0$ for $|\hb|$ small is partitioned  in stripes, ordered by 
increasing imaginary part,
 $$S_-^{m,0},~S_{m,0}^{m-1,0},~S_{m-1,0}^{m-1,1},~S_{m-1,1}^{m-2,1},\,....,\,S_{1,m-2}^{1,m-1},
~S_{1,m-1}^{0,m-1}, ~S_{0,m-1}^{0,m},~S_{0,m}^+\,,$$ 
respectively separated by the cuts in the same order,
 $$\gamma_{m,0},~\gamma_{m-1,0},~\gamma_{m-1,1},~\gamma_{m-2,1},\,....,\,\gamma_{1,m-1},
 ~\gamma_{0,m-1},~\gamma_{0,m},$$ 
 where a cut $\gamma_{j,k}$ is the arc of a suitable curve $\gamma_c$ from 
 $\hb_{j,k}$  to the origin, where it is tangent to the real axis.   The behavior of the function 
 $E_m(\hb)$ for $\hb\ra 0$ in the different  stripes is expressed in terms of the levels, 
 $$E_m^-(\hb),~E_0^+(\hb),~E_{m-1}^-(\hb),~E_1^+(\hb),\,...,\, 
 E_1^-(\hb)~E_{m-1}^+(\hb),~E_0^-(\hb),~E_m^+(\hb)\,,$$ 
 respectively in the same order of the 
 stripes. }
\bigskip

Thus, we have given a consistent picture of the minimal structure of the Riemann 
 surface of the level $E_m(\hb)$ free of cuts for $\Re\hb$ large, but 
 containing   the set of branch points and the corresponding cuts,
\be 
\{\hb_{j,k}\}_{(j,k)\in\mathbb{N}^2},\,\,\,\quad\,\,\,\{\gamma_{j,k}\}_{(j,k)\in\mathbb{N}^2}\,.\label{FSOS}\ee

\section{The  string, the  board and the sequence of the nodes}
In this final section we extend the analysis of the process of crossings expressed by the 
conjectures C1, C2, C2$'$, 
and Hypothesis H2.

In order to introduce the notion of string by a simple example, let us consider the 
\textit{harmonic oscillator} (\ref{HB2}), $ H(0)=p^2+x^2$. In this case we have four Stokes sectors 
in the complex plane: 
$$\Sigma_j=\{z\in\mathbb{C}\,,~|\arg(iz)-2j\pi/4|<\pi/4  \},\quad j=-1,0,1,2.$$ 
Given a 
level $E=E_n=(2n+1)$, $n\in\mathbb{N},$ with the corresponding  state $\psi_n,$ with a set of $n$ 
nodes $\{N_j\}$, $j=1,...,n$ and $n+1$  antinodes  $\{A_k\}$, $k=1,..,n+1$ on  the  string 
$\sigma=[I_-,I_+]$, where   $I_\pm=\pm\sqrt E$ are the turning points. In this case, the  string 
coincides  with the short Stokes line \cite{EGS}. 
As it is well known,  the  nodal sequence of the state $\psi_n$, naturally ordered as the real 
numbers, is the standard one, 
  $$S_n=(A_{1},~N_{1},~A_{2},...A_{n}~,N_{n},~A_{n+1})\in	\sigma.$$
In the Stokes complex, $X_2(E)$, there are also two anti-Strokes lines  $[\,I_+,+\infty)$,  
$(-\infty,I_-\,]$
representing the two   half-lines where the string is clamped. The  union of the string with these 
two semi-axes,
is the extended string, $\sigma^e=\mathbb{R}$. 
All the Stokes and anti-Stokes lines on $\mathbb{R}$ are exact, despite the fact that the Carlini 
corrections (\ref{ERCC}) are non vanishing. In this case, the  board is absent. 
Now, we go back to  the case of the \textit{cubic oscillator}. We have five Stokes 
sectors in the complex plane: 
$$\Sigma_j=\{z\in\mathbb{C}\,,~|\arg(iz)-2j\pi/5|<\pi/5  \},\,\,j=-2,-1,0,1,2.$$
We consider a  level $E=E^-_n(\hb)$, $n\in\mathbb{N},$  for a fixed $0<\hb<\hbar_n$. 
The corresponding state $\psi^-_n(\hb)$ is localized  about the single well  $x_-=-1/\sqrt{3}$.  	
The Stokes complex, $X_3(E)$, contains a subset   locally and topologically similar to the 	
harmonic complex $X_2(E)$, disconnected from the rest of $X_3(E)$. The   extended  string 	
$\sigma^e$ is a regular line going from the sector $S_{-1}$ to the sector  $S_1$. The nodes and 	
antinodes are in the string $\sigma$, the arc of the extended string  going from the point  	
$I_-$ to the point $I_0.$  The extended board $B^e$ is a regular line, and the board $B$ is a 	
half-line contained in $B^e$, going from $i\infty$ to the inversion point $I_+.$ Increasing 	
$\hb$ up to  $\hbar_n,$ the level crosses the other level $E^+_n(\hb)$ and $I_0$ comes in 	
contact 	with the board $B.$

For $\hb>0$ large, in the case of a positive level $E_m$, 	
$m\in\mathbb{N}$, the extended board $B^e$ is the imaginary axis, and the board is the semi axis 	
$B=[\,I_0,+i\infty).$ We call  continuation of the board the semi axis 	
$B^c=B^e-B$$=(-i\infty,I_0\,]$. The exceptional Stokes line is not only an approximation of the 	
board but coincides with it. The extended board coincides with the imaginary axis (Remark 3). 	

The extended string is always a line going from  the sector $S_{-1}$ to  the sector $S_1.$ The 
string is the arc of the extended string with end points $I_\pm.$
We have always a sequence of infinite zeros and of stationary points on the board $B.$ 

Let us fix $\hb>0$ small and consider the level $E=E_n^-(\hb)$ with the 
corresponding state $\psi=\psi_n^-(\hb)$, and its 
string $\sigma_n^-$ with the standard sequence of nodes
$$S_n^-=(A_{-n-1},~N_{-n}~,A_{-n},...A_{-2},~N_{-1},~A_{-1})\in	 \sigma_n^-.$$ 
On the board we have the 
standard sequence of zeros and stationary points $Z_j$, $Z'_j$ of the state 
$$Z_n^-=(Z'_0,~Z_0,~Z'_1,....)\in B_n^-.$$
 Also the other state $\psi_n^+$ has  the standard sequence of nodes
$$S_n^+=(A_{1},~N_{1},~A_{2},...A_{n},~N_{n},~A_{n+1})\in	 \sigma_n^+,$$ 
and the standard sequence of 
zeros 
$$Z_n^+=(Z'_0,~Z_0,~Z'_1,....)\in B_n^+.$$  
The sequences of nodes of the strings $S_n^-(\hb)$, 
$S_n^+(\hb)$ are stable and isolated from other zeros up to the crossing limit.

At the crossing, for $\hb=\hbar_n,$ we have the single level
 $$E^c_{n,n}=E_n^-(\hbar_n^-)=E_n^+(\hb^-_{n}),$$ 
 and the single state $\psi_{n,n}^c$ with the string $\sigma^c_{n,n}$
 given by the  union of the limits of the strings $\sigma_n^-,\sigma_n^+$.  At the limit, an arc of 
 the board of $\psi_n^-$ between the points $(I_+,I_0)$ becomes the string of $\psi_n^+$, and 
 analogously for the exchange of $+$ with  $-$. Thus, the exact short Stokes line becomes the union 
 of the two strings with the singular point $I_0.$ 
 Therefore the crossing state has the non standard  sequence of  nodes 
 \begin{eqnarray}
&{}& \!\!\!\!\!\!\!\!\!\!\!\!\!
S^c_{n,n}=(A_{-n-1},~N_{-n},~A_{-n},\,\,...\,\,A_{-2},~N_{-1},~A_{-1},~\,\cr
&{}& \phantom{XXXXXXX} A_{1},~N_{1},
~A_{2},\,\, ...\,\,A_{n},~N_{n},~A_{n+1})\in \sigma^c_{n,n},
 \end{eqnarray}
 and the  sequence of zeros on the board 
 $$Z^c_{n,n}=(Z_0,~Z'_0,~Z_1,....)\in B^c_{n,n}.$$ 
  Moreover,  for a parameter $h^p_{n}$,   close to $\hb_{n}$ from above, there is a
bilocalized state  $\psi^p_{2n+1}$, with the sequence of nodes:
\vfill\break
 \begin{eqnarray}
&{}& \!\!\!\!\!\!\!\!\!\!\!\!\!\!\!\! 
S^p_{2n+1}=(A_{-n-1},~N_{-n},~A_{-n},\,\,...,\,\,N_{-1},~A_{-1},~N_0=I_0,
\spazio{0.8}\cr
&{}&  \phantom{XXXXXXXXX} A_{1},~N_{1},~A_{2},\,\,...\,\,A_{n},~N_{n},~A_{n+1})\in\sigma^p_{2n+1},
 \end{eqnarray}
For a different parameter  $h^a_{n}$, near $\hb_{n}$, we have a state $\psi^a_{2n}$ with the 
sequence 
of nodes,
$$S^a_{2n}=(A_{-n},~N_{-n+1},...,N_{-1},~A_{-1}=I_0=A_{ 1},~N_1,..,N_{n},~A_{n+1})\in	
\sigma^a_{2n},$$
where  the antinodes $A_{\pm 1}$ are the limits of the stationary point $Z_0'$  and the antinode 
$A_0.$

Now we look at the crossings for complex $\hb.\,$ Let us fix $\,\hb\in\mathbb{C}^0,\,$ with 
$\,|\hb|\,$ 
small, and consider the level $\,E=E_j^-(\hb)$  with the corresponding state 
$\,\psi=\psi_j^-(\hb)\,$ and its 
string $\,\sigma_j^-\,$ with the standard sequence of nodes,
$$ S_j^-=(A_{-j-1},~N_{-j},~A_{-j},\,\,...\,\,A_{-2},~N_{-1},~A_{-1})\in	 \sigma_j^-.$$ 
On the board there is 
the standard sequence of zeros and stationary points $Z_l$, $Z'_l$ of the state: 
$$Z_j^-=(Z'_0,~Z_0,,Z'_1,\,\,....\,)\in B_j^-.$$
 Also the other state $\psi_k^+$ has  the standard sequence of nodes
$$S_k^+=(A_{1},~N_{1},~A_{2},\,\,...\,\,A_{k}~,N_{k},~A_{k+1})\in	 \sigma_k^+,$$ 
and the standard sequence of 
zeros 
$$Z_k^+=(Z'_0,~Z_0,~Z'_1,\,\,....\,)\in B_k^+.$$  
The sequence of nodes of the strings $S_j^-(\hb)$, 
$S_k^+(\hb)$ are stable and isolated from the other zeros up to the crossing in agreement with the 
semiclassical quantization condition and the analyticity.
$E^c_{j,k},$ 
 and the single state $\psi_{j,k}^c$.  
 The crossing state has the non standard  sequence of  nodes, 
\begin{eqnarray}
&{}& \!\!\!\!\!\!\!\!\!\!\!\!\!
S^c_{j,k}=(A_{-j-1},~N_{-j},~A_{-j},\,\,...\,\,A_{-2},~N_{-1},~A_{-1},~\,\cr
&{}& \phantom{XXXXX} A_{1},~N_{1},
~A_{2},\,\, ...\,\,A_{k},~N_{k},~A_{k+1})\in \sigma^c_{j,k},
\end{eqnarray}
while the  sequence of zeros on the board is 
$$Z^c_{j,k}=(Z_0,~Z'_0~,Z_1,\,\,....\,)\in B^c_{j,k}.$$ 
Moreover,  for a parameter $\hb^p_{j,k}$, close to $\hb_{j,k}$, there is a
bilocalized state  $\psi^p_{j+k+1}$, with the sequence of nodes:
\vfill\break
\begin{eqnarray}
&{}& \!\!\!\!\!\!\!\!\!\!\!\!\!\!\!\! 
S^p_{j+k+1}=(A_{-j-1},~N_{-j},~A_{-j},\,\,...,\,\,N_{-1},~A_{-1},~N_0,
\spazio{0.8}\cr
&{}&  \phantom{XXXXXXXXX} A_{1},~N_{1},~A_{2},\,\,...\,\,A_{k},~N_{k},~A_{kn+1})\in	
\sigma^p_{j+k+1},
\end{eqnarray}
where $N_0=I_0$ is  the limit of a zero $Z_0$ on $B.$
For a different parameter  $\hb^a_{j,k}$, near $\hb_{j,k}$, we have a state $\psi^a_{j+k}$ with the 
sequence of nodes,
$$S^a_{j+k}=(,A_{-j},~N_{-j+1},\,\,...\,\,N_{-1},~A_{-1}=A_{ 
1},~N_1,\,\,..\,\,N_{k},~A_{k+1})\in	
\sigma^a_{j+k},$$
where  one of the antinodes  $A_{\pm 1}=I_0$ is the limit of the antinode $A_0$, and the other is the limit of the stationary point 
$Z_0'$ as $\hb\ra\hb^a_{j,k}.$
\bigskip\bigskip
 
 \section{Appendix A. The Riccati equation and the semiclassical series expansion}
 \bigskip\bigskip
 
In order to define the exact Stokes complex and, in particular, the monochord consisting of the 
string and the board, we recall the Carlini semiclassical series expansion. 
We consider a Stokes sector of the complex plane far from the turning points and  we express two 
fundamental solutions of the Schr\"odinger equation, $$(\hb^2p^2+p_0^2(z))\psi(z)=0,\,\,\,\,\,p^2=\frac{d^2}{dz^2},\,\,\,\,p^2(z)=V(z)-E,$$ in the form, 
$$\psi(z)=\exp\Bigl(\frac 1\hb\int_0^zp_\hb(w)dw\Bigr),$$ 
where $p_\hb(z)$ satisfies 
the Riccati
equation, 
\be \phantom{XXX}p_\hb^2(z)+\hb p_\hb'(z)=p_0^2(z),\qquad p_0^2(z)=V(z)-E.\label{RE}\ee 
We solve formally by the  Carlini series $$p_\hb(z)\sim \sum_np_n(z)\hb^n,$$ where the coefficients 
are computed recursively starting from the two definitions of the classical momentum, 
$p_0(z)=\pm\sqrt {V(z)-E}$,
\vfill\break
\be  
&{}&\!\!\!\!\!\!\!\!\!\!\!\!\!\!\! p_1(z)=\frac{ip'_{0}(z)}{2p_0(z)},\cr
&{}&\!\!\!\!\!\!\!\!\!\!\!\!\!\!\! p_n(z)=-\frac{1}{2p_0(z)}\,\Bigl( 
\sum_{j=1}^{n-1}p_{n-j}(z)p_j(z)+ip'_{n-1}(z)\Bigr),\,\quad
 n=2,3,..\,\, .\label{ERCC}\ee
Defining $$p_\hb(z)=P_\chi(z)+\hb Q_\chi(z),\qquad\chi=\hb^2,$$ we get the  equation for 
$P_\chi(z)$, 
\be \,\,P_\chi^2(z)-p_0^2(z)=-\chi(Q_\chi^2(z)+Q_\chi'(z)),\quad\textrm{where} \quad 
Q_\chi(z)=-\frac{P'_\chi(z)}{2P_\chi(z)}.\label{SRE}\ee 
We thus have the equivalent expression of the solutions
 $$\psi(z)=\frac{1}{\sqrt{P_\chi(z)}}\exp\Bigl(\frac 1\hb\int_0^zP_\chi(w)dw\Bigr),$$     
where the Riccati solutions $ \pm P_\chi(z)$ have the even part of Carlini expansions 
 \be P_\chi(z)\sim 
 \sum_{j\in\mathbb{N}}\chi^{j}p_{2j}(z)\,\quad{\rm{with}}\,\,\,{\rm{truncations}}
 \quad\,\, P^N_\chi(z)\sim \sum_{j}^N\chi^{j}p_{2j}(z)
 .\label{ESM}\ee 
Actually, we associate to the 
asymptotic expansion the continued fraction, as its formal sum, \be P^{cc}_\chi(z)=p_0(z)+\chi 
p_2(z)/(1-\chi (p_4(z)/p_2(z))/..)\label{IER}\ee
                    In particular, we have the Pad\'e $[1,1]$ of the series (\ref{ESM}),  
                    $$P^{[1,1]}_\chi(z)=p_0(z)+\chi p_2(z)/(1-\chi (p_4(z)/p_2(z))),$$ where
 $$p_2(z)=\frac{q_0^2(z)+q'_0(z)}{2p_0(z)},\,\,\quad
 \,\,\,p_4(z)=\frac{(2q_0(z)q_2(z)+q_2'(z))-p_2^2(z)}{2p_0(z)}$$ 
 and  
 $$q_0(z)=\frac{p'_0(z)}{2p_0(z)}=\frac{V'(z)}{4p_0^2(z)},\qquad\,\,q_2(z)=
 \frac{p_2'(z)}{2p_0(z)}-\frac{p_0'(z)p_2(z)}{2p_0^2(z)},$$
where $q_{2n}=p_{2n+1}.$

At the limit of the turning points, the coefficients of the Carlini expansion are singular, but the 
diagonal Pad\'e approximants are regular.
\medskip

We recall that each Stokes line is  defined by a turning 
point as starting point and the condition on the  
field of directions $dz$ given by, 
$$p_0^2(z)dz^2<0.$$ 
The exact Stokes and anti-Stokes lines are defined by the turning point as starting point and the 
conditions 
\be P_\chi^2(z)dz^2 < 0, \label{EXACTSTOKES}\ee
on the field of directions respectively.
At a given approximation we treat separately  a neighborhood of the each turning point, linearizing 
the potential  and approximating the beginning of the Stokes lines by the Airy solutions. This 
neighborhood should shrink to a point in the limit of exact approximation.
\bigskip

\textsc{Remark 4}
Let us consider the Hamiltonian (\ref{MHB}). A positive level $E_m(\hb)$, corresponds to a negative 
eigenvalue $E=-E_m(\hb)$ of (\ref{MHB}). The imaginary axis appears as the real axis in the 
representation (\ref{MHB}) and coincides with the extended board. And the board coincides with the 
exceptional Stokes line. Actually, in the representation (\ref{MHB}), we have the reality of all 
$p_0(y)$, and $P^N_\chi(y)$ for all $N\in\mathbb{N},$ so that the board is on the imaginary axis. 
\bigskip

\section{Appendix B. Numerical aspects.}

\bigskip
 
We discuss some numerical results about the instability of the Stokes complex at the critical 
energy. In particular, for  $\hb=0$, we consider the 
monochord consisting of the  string (the short Stokes line) and the  board (the exceptional 
Stokes line). For $\hb>0$ small, we consider the 
approximate monochord consisting of the approximate string (the short Stokes line) and the 
approximate board (the exceptional Stokes line). In case of positive energy and positive parameter, 
the approximate 
board is actually exact. 
The short Stokes lines computed are good approximations of the 
corresponding strings because the nodes and antinodes appears to lie in it. The results are in 
agreement with the conjectures C1, C2.

\begin{table}[h]
	\begin{small}
		\begin{center}
			\begin{tabular}{c|c|c|c}
				n & {$\hbar_n^p$}  & {$E_n^p$}\\
				\hline
				3 & 0.0558&  0.35200  \\
				4 & 0.0438&  0.35209  \\
				5 & 0.0306 &  0.35218  \\
				6 & 0.0236 &  0.35221 \\
				7& 0.0130 &  0.35223 \\
				\hline
			\end{tabular}
			\qquad\qquad\qquad
			\begin{tabular}{c|c|c|c}
				n & {$\hbar_n^a$}  & {$E_n^a$}\\
				\hline
				3 & 0.0615&  0.35317  \\
				4 & 0.0473&  0.35287  \\
				5 & 0.0323 &  0.35261  \\
				6 & 0.0247 &  0.35244 \\
				7& 0.0133 &  0.35235 \\
				\hline
			\end{tabular}						
		\end{center}
	\end{small}
	\caption{ The values of  $\hbar_n^p$, $E_n^p$ (left) and
		the values of  $\hbar_n^p$, $E_n^p$ (right) for   $3\leq n\leq 7$ }
\end{table}

Consider first some facts occurring at  $\hb=0$.
The energy $E^c=0,352268..$ is a critical point of the monochord.
When $E-E^c>0$ is small, the string is a regular arc of a curve (Fig. 2) separated from the 
board.
Small variation of the energy $E^c$ on the complex plane can yield the separation of 
one half of the string, which becomes the new string, where the other half string remains attached 
and becomes an extension of the board (Fig. 3).
For $E^\pm_n(0)=\mp i2/3\sqrt{3}$, $\forall n\in\mathbb{N},$
the strings are the points $I_\pm=I_0=x_\pm=\pm1/\sqrt{3},$ respectively (Fig. 4). \\
For $\hb>\hbar_n$ and positive energy $E=E_m(\hb)$, $[m/2]=n$,  the  board $B(E)$ is a half-line on 
the 
imaginary axis  and the  string is $P_x$-symmetric. In the semiclassical limit, for a diverging 
$m(\hb)$ 
such that 
\be E_{m(\hb)}(\hb)\ra E\quad {\mathrm{as}}\quad \hb\ra 0^+ \label{LSG}\ee 
the string at the energy $E$ is 
the semiclassical localization of the state $\psi_{m(\hb)}(\hb)$ at the same limit (\ref{LSG}). 
 
In Fig. 5  the trajectory of the spectrum about the crossing at $\hb_3$ is represented. The complex 
levels are given by $\Re E+\Im E.$ 

In Fig. 6 we show the nodes and antinodes of the state $\psi_{2n}(z)$ with energy $E_{2n}$ at a  
fixed $\hb>h^a_n$. In the same figure we also add an imaginary zero $Z_0$ and an imaginary 
stationary point $Z'_0$ of $\psi_{2n}(z)$ in the board $B(E_{2n})$. 

Fig. 7 illustrates the nodes and antinodes of the state $\psi_{2n}(z)$ with energy $E_{2n}$ at a  
parameter $\hb>h^a_n$. In the same figure we also show an imaginary zero $Z_0$ and an imaginary 
stationary point $Z'_0$ of $\psi_{2n}(z)$ on the board $B(E_{2n})$.

At  the value $\hb=h_n^a$ of the parameter   
the  imaginary  antinode $A_0$ of the state $\psi_{2n}(\hb)$ coincides with the 
turning point $I_0$ and the stationary point $Z'_1$. At  a parameter $\hb>h_n,$  $\hb<h_n^a$,  the 
sequence of nodes of $\psi_{2n}(\hb)$ is the same as the critical state at the crossing 
$\psi_{n,n}^c$. See the Fig. 8 for $n=3$. 

For $\hb_n<\hb<\hb_n^p$ the sequence of nodes of $\psi_{2n+1}(\hb)$ is the 
same as the critical state at the crossing $\psi_{n,n}^c$.  See the Fig. 9 for $n=3$. 

For $0<\hb<\hb_n,$ the $n$ nodes of $\psi_n^-(\hb)$ are near the approximate string, 
Fig. 10.

 Near the crossing, for $|\hb-h_n|$ small, the part of the string near the turning point 
$I_0$ is difficult to follow numerically. The reason is in the breaking symmetry at $\hb=\hb_n.$ We 
have disregarded this part of  the string in  the Figures $7,\,8,\,9$.

At   $\hb=\hbar_n^p$, the  imaginary  node of the state $\psi_{2n+1}(\hb)$ 
coincides with the imaginary turning point, $N_0=I_0$, while at   $\hb=\hbar_n^a$ the  
imaginary  antinode of the state $\psi_{2n+1}(\hb)$ coincides with the turning point, $A_0=I_0$.

\bigskip

\textsc{Aknowlodgements.} It is a pleasure to thanks Professor Andr\'e Martinez for many long and 
useful discussions at the beginning of this research. We thanks also  C. Giberti and C. Vernia for 
useful suggestions about the numerical methods.

\end {document}